\documentclass[conference,compsoc]{IEEEtran}
\usepackage[nospace]{cite}
\ifCLASSINFOpdf
\else
\fi

\usepackage{url}
\usepackage{graphicx}
\usepackage{tabularx}
\usepackage{algorithm}
\usepackage[noend]{algpseudocode}
\usepackage{listings}
\usepackage{comment}
\usepackage{pifont}
\usepackage{subfig}
\usepackage{color}
\usepackage{xcolor}
\usepackage{array}
\usepackage{tabularx}
\usepackage{amsfonts}
\usepackage{amsthm}
\usepackage{multirow}
\usepackage{amsmath}
\usepackage{amssymb}
\usepackage{longfbox}
\newtheorem{theorem}{Theorem}
\newtheorem{definition}{Definition}
\usepackage{lipsum}  
\usepackage{booktabs} 
\usepackage{enumitem}
\usepackage{hyperref}
\usepackage{colortbl}
\usepackage{array}

\hypersetup{
    colorlinks=true,       
    linkcolor=teal,        
    citecolor=teal,        
    urlcolor=teal          
}

\newcommand{\ToolName}{\textit{Empc}}
\newcommand{\SearchToolStyle}[1]{\textsf{\small #1}}
\newcommand{\SearchToolSmallStyle}[1]{\textsf{\footnotesize #1}}
\newcommand{\ProgramStyle}[1]{\textrm{#1}}
\newcommand{\CommandStyle}[1]{\textsf{\footnotesize #1}}
\newcommand{\CodeStyle}[1]{\textsf{\small #1}}
\newcommand{\AlgComment}[1]{\Comment{\textcolor{purple}{#1}}}

\newcommand{\ReduCellColor}[1]{\cellcolor{green!25} {#1}}
\newcommand{\ToolCellColor}[1]{\cellcolor{blue!20} {#1}}
\newcommand{\CompCellColor}[1]{\cellcolor{red!30} {#1}}
\newcommand{\revise}[1]{#1}

\begin{document}
%
\title{\ToolName{}: Effective Path Prioritization for Symbolic Execution with Path Cover}

%
\author{\IEEEauthorblockN{Shuangjie Yao, Dongdong She}
\IEEEauthorblockA{Hong Kong University of Science and Technology\\
\{syaoap, dongdong\}@cse.ust.hk}}





\maketitle

\begin{abstract}
Symbolic execution is a powerful program analysis technique that can formally reason the correctness of program behaviors and detect software bugs. It can systematically explore the execution paths of the tested program. But it suffers from an inherent limitation: path explosion. Path explosion occurs when symbolic execution encounters an overwhelming number (exponential to the program size) of paths that need to be symbolically reasoned. It severely impacts the scalability and performance of symbolic execution.   

To tackle this problem, previous works leverage various heuristics to prioritize paths for symbolic execution. They rank the exponential number of paths using static rules or heuristics and explore the paths with the highest rank. However, in practice, these works often fail to generalize to diverse programs. 

In this work, we propose a novel and effective path prioritization technique with path cover, named \ToolName{}. Our key insight is that not all paths need to be symbolically reasoned. Unlike traditional path prioritization, our approach leverages a small subset of paths as a minimum path cover (MPC) that can cover all code regions of the tested programs. To encourage diversity in path prioritization, we compute multiple MPCs. We then guide the search for symbolic execution on the small number of paths inside multiple MPCs rather than the exponential number of paths. 

    
We implement our technique \ToolName{} based on KLEE. We conduct a comprehensive evaluation of \ToolName{} to investigate its performance in code coverage, bug findings, and runtime overhead. The evaluation shows that \ToolName{} can cover 19.6\% more basic blocks than KLEE's best search strategy and 24.4\% more lines compared to the state-of-the-art work \SearchToolStyle{cgs}. \ToolName{} also finds 24 more security violations than KLEE's best search strategy. Meanwhile, \ToolName{} can significantly reduce the memory usage of KLEE by up to 93.5\% and reduce the number of symbolic states by up to 88.6\%.
\end{abstract}


%


\section{Introduction} \label{sec:intro}
Symbolic execution\cite{king1975new, boyer1975select, king1976symbolic, howden1977symbolic, cadar2013symbolic, baldoni2018survey} is a powerful program analysis technique that has been widely used in software testing \cite{poeplau2020symbolic, poeplau2021symqemu, godefroid2005dart, sen2005cute, stephens2016driller, kukucka2022confetti, yun2018qsym}, formal verification\cite{chau2017symcerts, hernandez2017firmusb, chau2019analyzing, susag2022symbolic}, and automated reasoning\cite{schwerhoff2016advancing, coppa2017rethinking, hu2020automated, kuts2021towards}. 
Unlike conventional software testing approaches that execute the program with concrete input values, symbolic execution treats inputs as symbolic variables, allowing it to explore multiple execution paths simultaneously\cite{cadar2013symbolic, baldoni2018survey}. This method systematically generates inputs that drive the program to various states, helping to uncover hidden bugs and security vulnerabilities by checking against a set of correctness specifications\cite{baldoni2018survey, li2013software, ruaro2021syml}. By using symbolic values rather than actual data, it provides a comprehensive analysis of the program's behavior, enabling testers and developers to verify the correctness and robustness of complex software systems. As such, symbolic execution plays a crucial role in improving software reliability and security, making it an essential topic of study not only in academic research but also in industrial practice such as Microsoft\cite{godefroid2008automated}, IBM\cite{artzi2008finding}, NASA\cite{puasuareanu2010symbolic, mirzaei2012testing}, Baidu\cite{baidu2018concfuzzer} and so on.

\vspace{0.1cm}
\noindent
\textbf{Bottleneck.} One of the main bottlenecks of symbolic execution is the path explosion\cite{anand2008demand, cadar2013symbolic, baldoni2018survey}. This is caused by the characteristic that a symbolic execution engine forks off the state at every branch of the program. Hence, each conditional statement in the program can potentially double the number of paths, leading to an exponential number of paths to explore in symbolic execution. The path explosion not only incurs prohibitive computation costs on CPU and memory but also severely limits the scalability of symbolic execution. In practice, symbolic execution often fails to reason large real-world programs.

\vspace{0.1cm}
\noindent
\textbf{Existing work.} Modern symbolic execution engines prioritize promising paths using various search strategies to tackle the path explosion problem\cite{baldoni2018survey}.
KLEE\cite{cadar2008klee}, one of the most popular symbolic execution engines, incorporates multiple search strategies including \SearchToolStyle{bfs} (breadth-first search), \SearchToolStyle{rps} (random-path search) and \SearchToolStyle{nurs} (non-uniform random search). 
\SearchToolStyle{md2u}~\cite{burnim2008heuristics} leverages the control-flow graph to steer the search towards the closest uncovered branches. \SearchToolStyle{sgs} (subpath-guided search)\cite{li2013steering} prioritizes the least explored subpaths. Kapus et al. propose an approach aimed at exploring pending paths already known to be feasible\cite{kapus2020pending}. \textit{Ferry}\cite{zhou2022ferry} uses the dependence of the focused variable to guide the search. A recent work \SearchToolStyle{cgs} (concrete-constraint guided search)\cite{sun2024concrete} favors uncovered branches with concrete variable assignments. Although these static heuristics improve the search efficiency of symbolic execution, they cannot be generalized to diverse programs.

\vspace{0.1cm}
\noindent
\textbf{\ToolName{}.} 
In this work, we introduce a novel approach called \ToolName{} to mitigate the path explosion problem using path cover. Unlike prior works that rank the exponential number of paths following some static heuristics, we leverage a small subset of paths as a minimum path cover (MPC) such that it can cover all the code regions of the program using the minimum number of paths. We further compute multiple MPCs to ensure diverse choices in path prioritization. We then search over the small subset of paths rather than the entire exponential number of paths. The key insight of our work is that not all paths need to be symbolically solved if we want to cover all the code regions of the program. We model the path prioritization in symbolic execution as an MPC problem~\cite{bang2008digraphs} in the graph theory domain and guide the search of symbolic execution over path cover.      

However, there are two practical challenges when applying our path-cover-based search in symbolic execution. Firstly, the inter-procedural control-flow graphs (iCFG) of programs are complex graphs with many cycles such as caller-callee cycles and loop cycles, while the existing MPC algorithm can only work on directed acyclic graphs (DAG). Hence, some approximation algorithms to eliminate different types of cycles in iCFG is needed.
Second, some paths in our small set of path cover can be infeasible during symbolic execution. It happens when some path constraints are proved to be infeasible by the SMT solver in the symbolic execution engine.

We propose \ToolName{} to solve these challenges. Our approach includes two main modules: path prioritization via multiple MPCs and infeasible path handling. We first propose some approximation algorithms for the original MPC algorithm, performing graph transformation on the vanilla iCFG to obtain many acyclic iCFG subgraphs. The goal of such an iCFG graph transformation is to enable trackable MPC computation. Otherwise, it will be an NP-hard problem to compute MPCs on a graph with cycles. We then compute multiple MPCs on the transformed iCFG to increase the path diversity. And the small number of paths in multiple MPCs can cover  
all nodes on the iCFG. At run-time of symbolic execution, \ToolName{} compares and finds a solvable path in multiple MPCs to execute. When \ToolName{} encounters an infeasible path that the SMT solver cannot solve, the infeasible path handling module will be invoked and it will discover a new path using the program dependence information.

\vspace{0.1cm}
\noindent
\textbf{Evaluation.} We implement \ToolName{} as a searcher module on top of KLEE. To investigate the performance of \ToolName{}, we conduct a comprehensive experiment on 12 real-world programs. Our evaluation shows that \ToolName{} increases the basic block coverage by 19.6\% compared to KLEE's best search strategy and the line coverage by 24.4\% compared to the state-of-the-art work \SearchToolStyle{cgs}~\cite{burnim2008heuristics} in arithmetic mean on these real-world programs. Meanwhile, \ToolName{} significantly reduces KLEE's memory cost by up to 93.5\% and further cuts the number of execution states in the symbolic execution engine by up to 88.6\%. Our evaluation shows that the runtime overhead of \ToolName{} is minimal, with an average of 12\% on 12 programs. In the end, we show that \ToolName{} finds 24 more security violations than KLEE's best search strategy.

\noindent
\textbf{Contributions.} Our main contributions are as follows.
\begin{itemize}
    \item We model path prioritization in symbolic execution as a classic path cover problem in graph theory domain. 
    \item We propose a novel search strategy for symbolic execution by searching over the small path cover instead of the exponential number of all possible paths. 
    \item We implement our technique as a prototype \ToolName{} on top of KLEE and open source our tool \revise{at Github (\href{https://github.com/joshuay2022/empc}{https://github.com/joshuay2022/empc})} to foster further research in this domain.
    \item We perform a comprehensive evaluation of \ToolName{} to investigate its performance in code coverage and bug finding. Our result shows that \ToolName{} can achieve 19.6\% more basic block coverage and 24.4\% more line coverage over the state-of-the-art search strategies.  
\end{itemize}


\section{Path Prioritization as Path Cover Problem}
\begin{figure*}[!t]
\centering

\centering
\subfloat[Sample code.]{\includegraphics[width=0.24\linewidth]{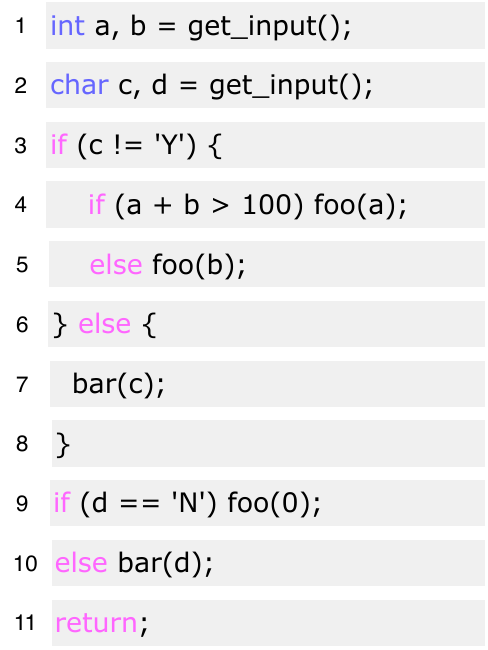}\label{subfig:sample_code}}
  \hspace{0.1cm}
\subfloat[Program CFG.]{\includegraphics[width=0.22\linewidth]{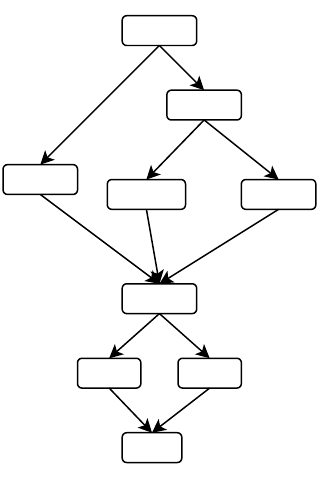}\label{subfig:sample_cfg}}
 \hspace{0.1cm}
\subfloat[Path explosion.]{\includegraphics[width=0.26\linewidth]{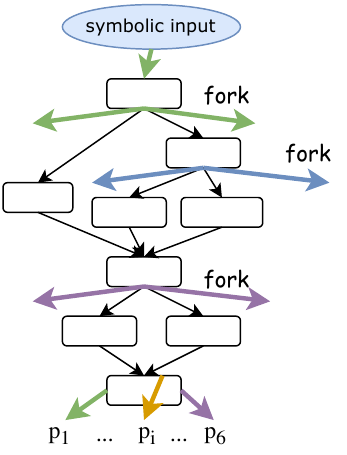}\label{subfig:sample_exp}}
 \hspace{0.1cm}
\subfloat[A minimum path cover.]{\includegraphics[width=0.20\linewidth]{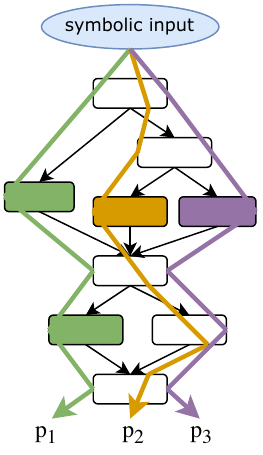}\label{subfig:sample_paths}}
\caption{\textbf{\small  This figure shows a motivating example including sample code, its CFG, path explosion problem and our minimum path cover method. Our method only uses 3 paths to cover all basic blocks of this program with 3 branches while the number of paths $|P|=6$ if there is a path explosion problem. }}
\label{fig:motivation}
\end{figure*}

In this section, we first introduce some background knowledge for the MPC problem, then we formulate the path prioritization in symbolic execution as a path cover problem.

\subsection{Minimum Path Cover} \label{sec:mpc}
Given a graph $G = (E, V)$, a path cover for all vertices $V$ is defined as a set of paths $P_C=\{p_1, p_2, ..., p_k\}$ such that for each $v_i \in V$, there is at least one path $p_j$ covering node $v_i$\cite{bang2008digraphs, ntafos1979path}. For a given graph, there could exist multiple path covers. An MPC $P_m$ is a path cover whose size is minimum among all possible path covers\cite{bang2008digraphs, ntafos1979path}. Computing MPC on general graphs is an NP-hard problem. However, researchers propose multiple algorithms to solve it in polynomial time on directed acyclic graphs (DAG)\cite{dilworth1987decomposition, fulkerson1956note, ntafos1979path}. In this work, we use Ntafos's maximum matching method\cite{ntafos1979path} to compute the MPC. We show this algorithm in Algorithm \ref{alg:sin_mpc_comp}.

\begin{algorithm}
\footnotesize
\caption{Compute One Minimum Path Cover} 
\label{alg:sin_mpc_comp} 
\lstset{basicstyle=\ttfamily\footnotesize, breaklines=true}
\begin{tabular}{|lp{2.7in}|}\hline
\textbf{Input}:
    & A directed acyclic graph $G(V,E)$ \\
\hline
\end{tabular}
\begin{algorithmic}[1]

\State $min\_path\_cover \gets \{\}$
\State $n \gets |V|$, $m \gets |E|$
\State $V_b \gets \{x_1,x_2,...,x_n\}\cup \{y_1,y_2,...,y_n\}$
\State $E \gets \{(x_i,y_i)|1\leq i\leq n, v_i ~\mathrm{reaches} ~v_j ~\mathrm{in} ~G\}$
\State A bipartite graph $G_b \gets (V_b, E_b)$
\State A maximum matching $M_m \gets \mathrm{Hopcroft\_Karp\_Algorithm(}G_b\mathrm{)}$, $M_m \gets \{(x_{i1}, y_{j1}),(x_{i2}, y_{j2}),...,(x_{ic}, y_{jc})\}$
\For{$(x_{ik},y_{jk}) \in M_m$}
    \If{$(v_{ik}, v_{jk}) \notin E$}
        \State $p_{sub} \gets v_{ik}(v_{ik}, v'_{ik})...v_{jk}$ \AlgComment{construct a subpath}
    \Else
        \State $p_{sub} \gets v_{ik}(v_{ik}, v_{jk})v_{jk}$ \AlgComment{construct a subpath directly}
    \EndIf
    
    \For{TRUE}
        \State find a path $p_i$ from $min\_path\_cover$ with a same end-vertex as $p_{sub}$
        \If{$p_i$ doesn't exist}
            \State $min\_path\_cover \gets min\_path\_cover ~\cup~ \{p_{sub}\}$
            \State \textbf{break}
        \Else
            \State merge $p_i$ and $p_{sub}$ into $p'_i$
            \State $min\_path\_cover \gets min\_path\_cover \backslash \{p_i\}$
            \State $min\_path\_cover \gets min\_path\_cover \cup \{p'_i\}$
            \State $p_{sub} \gets p'_i$
        \EndIf
    \EndFor
\EndFor

\State
\Return $min\_path\_cover$

\end{algorithmic}
\end{algorithm}

\subsection{Problem Formulation}
We formulate path prioritization in symbolic execution as a classic path cover problem. We denote the iCFG of the tested program as $G=(E, V)$. Since Algorithm \ref{alg:sin_mpc_comp} is only applied to DAGs, we propose some approximation algorithms including graph transformation in Section~\ref{sec:graph} to compute MPCs. There are a total exponential number of paths to be symbolically reasoned on the graph $G$, denoted as $P_G = \{p_1, p_2,..., p_k\}$. We then compute an MPC $P_m = \{p_i,p_j,...,p_n\}$ such that all vertices in $V$ can be covered by the $n$ path in $P_m$ and the size of $P_m$ is much lower than $P_G$, that is, $|P_m| \ll |P_G|$. We then guide symbolic execution to search on the path cover $P_m$.

\section{Overview}

\subsection{Motivating Example}

In this section, we give a motivating example for our work. Figure \ref{fig:motivation} is a sample program and its CFG is shown in Figure \ref{subfig:sample_cfg}. This sample program shows a common case in real-world programs. Normally, the common symbolic execution engine forks off a state at each branch, so there will be 6 paths in this program, as shown in Figure \ref{subfig:sample_exp}. This is because the first two forks only yield 3 subpaths and all 3 subpaths then merge at the last fork, and finally yield a total of $3\times 2=6$ paths. If we are aimed at maximizing code coverage, it is obvious that at least 5 complete paths are needed to cover all basic blocks in this program in the worst case. However, to cover all basic blocks in CFG, our MPC method uses only 3 complete paths $p_1$, $p_2$ and $p_3$ as shown in Figure \ref{subfig:sample_paths}. Because we have a predefined path cover, we only need to select the forked state that matches the path in the MPC at each branch. Ultimately, we use a small subset of paths to cover all basic blocks by ignoring most states.

\subsection{Workflow}

\begin{figure}[!t]
\centering
\includegraphics[scale=0.58]{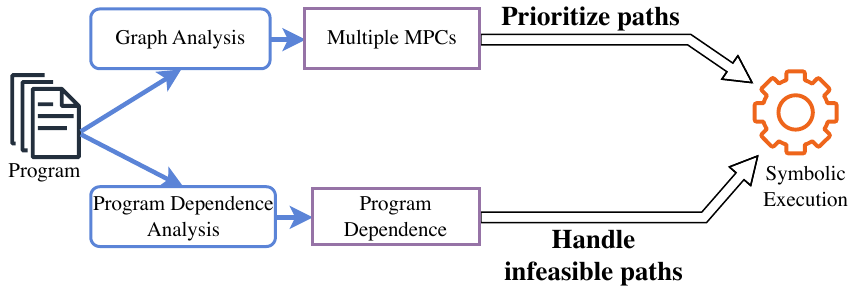}
\caption{\textbf{\small Workflow of \ToolName{}.}}
\label{fig:workflow}
\end{figure}


Figure \ref{fig:workflow} shows the complete workflow of our symbolic execution framework. \ToolName{} first performs a graph analysis on the program to generate multiple MPCs. Then, it uses multiple MPCs to prioritize a subset of paths to guide symbolic execution at run-time. Moreover, \ToolName{} leverages a simple program dependence analysis to capture the program dependence between a branch and its dependent basic blocks. When symbolic execution encounters an infeasible path after searching multiple MPCs, \ToolName{} will find a new state to continue execution using the dependence information. 

\section{Methodology}

\revise{In this section, we introduce \ToolName{} in detail. \ToolName{} consists of two components: path prioritization via multiple MPCs and infeasible path handling. 
}

\begin{figure*}[!t]
\centering

\subfloat[Original iCFG.]{\includegraphics[width=0.27\linewidth]{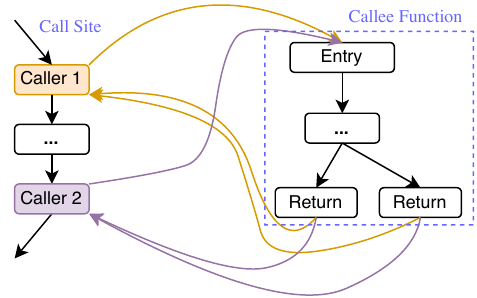}\label{subfig:func_trans_ori}}
  \hspace{0.5cm}
\subfloat[Add a virtual return node.]{\includegraphics[width=0.27\linewidth]{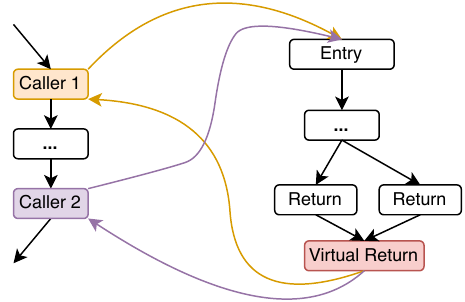}\label{subfig:func_trans_vnode}}
 \hspace{0.5cm}
\subfloat[Graph split.]{\includegraphics[width=0.27\linewidth]{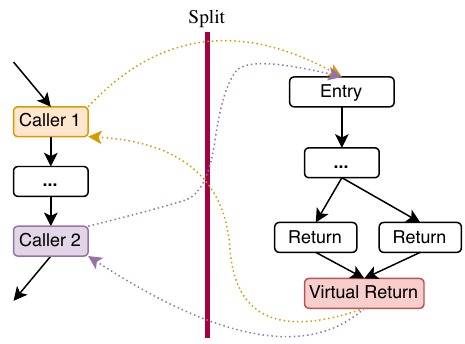}\label{subfig:func_trans_split}}
\caption{\textbf{\small Transform an iCFG with caller-callee cycles into two subgraphs. The original iCFG contains two callers and the edges in yellow and purple are calling and returning edges in Figure \ref{subfig:func_trans_ori}. We transform the iCFG by adding a virtual return node in red in Figure \ref{subfig:func_trans_vnode}. We remove the calling and returning edges for graph split and mark them as dotted lines in Figure \ref{subfig:func_trans_split}. The right subgraph is a one-entry-one-exit subgraph.}}
\label{fig:func_trans}
\end{figure*}

%
%
\subsection{Path Prioritization via Multiple MPCs}
We explain in detail how to compute multiple MPCs in iCFG and prioritize the paths for symbolic execution. Firstly, we transform the vanilla iCFG with cycles into many acyclic iCFG subgraphs. For each acyclic iCFG subgraph, we enumerate all MPCs using the maximum matching method in graph theory. Multiple MPCs can provide diverse path selection in the path prioritization phase of symbolic execution. In the end, we guide the symbolic execution to search a subset of paths instead of all possible paths with the help of multiple MPCs.
\subsubsection{MPC Computation in Transformed iCFG} \label{sec:mpc_construct}
\label{sec:graph}
\revise{
We start with the computation of a single MPC in an iCFG of a program. Computing an MPC in an iCFG is challenging because the iCFG contains cycles but computing an MPC in such a cyclic graph is an NP-hard problem\cite{ntafos1979path}. In graph theory, there are only polynomial run-time algorithms in the directed acyclic graph (DAG) as shown in Algorithm \ref{alg:sin_mpc_comp}. Therefore, we try to find a solution that closely approximates the MPC in iCFG via approximation algorithms, which are typical solutions to NP-hard problems\cite{williamson2011design}. Generally, our approximation algorithms transform the cyclic iCFG into DAGs and then compute MPC in the transformed iCFG. Note that we \emph{only} transform the iCFG in the graph analysis step. The path exploration including path prioritization at run-time will not be influenced by our graph transformation.}

\begin{figure*}[!t]
\centering

\subfloat[Original iCFG.]{\includegraphics[width=0.18\linewidth]{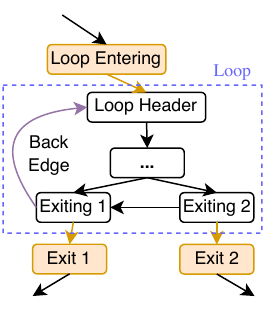}\label{subfig:loop_trans_ori}}
  \hspace{0.15cm}
\subfloat[Add virtual nodes and remove the back edge.]{\includegraphics[width=0.38\linewidth]{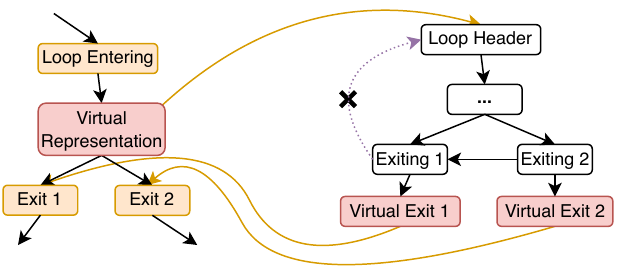}\label{subfig:loop_trans_vnode}}
 \hspace{0.15cm}
\subfloat[Graph split.]{\includegraphics[width=0.38\linewidth]{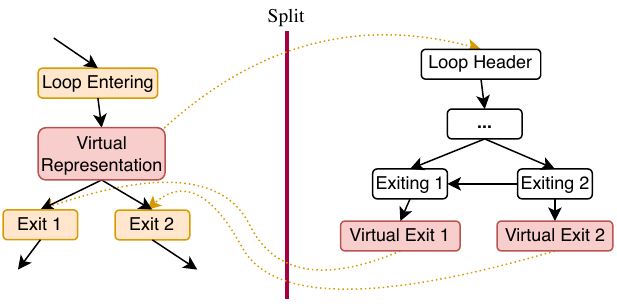}\label{subfig:loop_trans_split}}
\caption{\textbf{\small Transform the iCFG with loop cycles into two subgraphs without cycles. The original iCFG contains a loop with a loop body and a back edge in Figure \ref{subfig:loop_trans_ori}. The iCFG is transformed by removing back edges and adding several virtual nodes in red in Figure \ref{subfig:loop_trans_vnode}. The loop body is represented by a virtual node in the transformed iCFG and the virtual exit nodes match up with the exit nodes respectively. The loop entering and exiting edges are removed for graph split and marked as dotted lines in Figure \ref{subfig:loop_trans_split} and the right subgraph is a loop subgraph.}}
\label{fig:loop_trans}
\end{figure*}

\revise{
The iCFG includes two types of cycles: \textbf{caller-callee cycles} and \textbf{loop cycles}. Figure~\ref{subfig:func_trans_ori} provides a simple example of a caller-callee cycle. This cycle begins at the node \CodeStyle{Entry} in the callee function, traverses the entire callee function, and then returns to the node \CodeStyle{Caller 1}. Subsequently, \CodeStyle{Caller 1} transitions to \CodeStyle{Caller 2}, which then returns to \CodeStyle{Entry} in the callee function. Thus, a caller-callee cycle typically occurs when a callee function is invoked by multiple callers. Additionally, as illustrated in Figure~\ref{subfig:loop_trans_ori}, a \textbf{loop cycle} starts at the \CodeStyle{Loop Header}, traverses through the loop body, and returns to the \CodeStyle{Loop Header} via a back edge.
}

\revise{Both types of cycles—caller-callee and loop cycles—comprise the following components, as defined by \ToolName{}. A \textbf{cycle body} is a set of vertices and edges that form the core structure of the cycle. \textbf{Cycle edges} are back edges that point to an ancestor of the current node, creating the cycle. \textbf{Connecting edges} are edges that connect the cycle body to other parts of the graph or within the cycle itself. As shown in Figure~\ref{subfig:func_trans_ori}, in a \textbf{caller-callee cycle}, the cycle body corresponds to the function itself, which is enclosed in the blue dashed box. The connecting edges include the calling and returning edges of the first caller, which are highlighted as yellow edges. The cycle edges are the calling and returning edges of the second caller, highlighted as purple edges. Similarly, as shown in Figure~\ref{subfig:loop_trans_ori}, in a \textbf{loop cycle}, the cycle body refers to the loop body, excluding the back edges, and is enclosed in the blue dashed box. The connecting edges consist of entering and exiting edges, highlighted as yellow edges. The cycle edges are the back edges, highlighted as purple edges.}

\revise{Based on this analysis, we approximate the computation of the MPC in an iCFG through the following two steps. Firstly, we transform the iCFG into an acyclic graph by removing cycles edges in the graph analysis phase, which precedes path prioritization, because this operation enables the computation of MPCs using Algorithm \ref{alg:sin_mpc_comp}. Thus, the transformed iCFG includes the cycle body but excludes the cycle edges, ensuring that the vertices within the cycle body appear at most once on a path. Consequently, our approximation simplifies the computation on the original directed cyclic graph by reducing it to a directed acyclic graph, effectively ignoring the cycle edges. Secondly, we analyze each cycle body independently from the transformed iCFG as a transformed subgraph. We split the transformed graph by removing the connecting edges. This step is important because the cycles need to be reintroduced after computation to ensure program completeness. The subgraph contains subpaths of the paths in the iCFG, allowing us to compute the MPC within the transformed subgraph. During path prioritization, \ToolName{} can still account for repeated executions of the cycle body by considering a new path from the MPC of the cycle body each time it repeats. Finally, we propose new theorems and approaches to make the computation of MPCs in each subgraph provably correct. Based on these analyses, we propose detailed approaches for transforming each of the two types of cycles described above.}

\vspace{0.1cm}
\noindent
\textbf{Transform Caller-Callee Cycles.} Based on the several steps described above, \ToolName{} first adds a virtual node as the successor of all return nodes in the function body (see Figure~\ref{subfig:func_trans_vnode}), and then splits the graph into two parts, as shown in Figure~\ref{subfig:func_trans_split}. The left part represents the transformed graph, while the right part is a subgraph with one entry and one exit node. We classify such subgraphs as one-entry-one-exit graphs. As the name suggests, each one-entry-one-exit graph $G_{sub}(V_{sub}, E_{sub})$ has exactly one entry vertex $v_{ss}$ and one exit vertex $v_{st}$. Its formal definition is provided in Definition~\ref{def:oeoe_subg}. The right part of Figure~\ref{subfig:func_trans_split} illustrates a one-entry-one-exit subgraph. These subgraphs can be generated using Algorithm~\ref{alg:split_oeoe}. For function calls, the entry block and exit block of a function can be easily identified, enabling the generation of a one-entry-one-exit subgraph via Algorithm~\ref{alg:split_oeoe}.

\begin{definition}[One-Entry-One-Exit Subgraph] \label{def:oeoe_subg}
    A subgraph $G_{sub}(V_{sub}, E_{sub})$ is a one-entry-one-exit subgraph of $G(V,E)$ iff. i) $G_{sub}$ is an induced subgraph of $G$; ii) $G_{sub}$ has two vertices $v_{ss}$ and $v_{st}$ where in-degree of $v_{ss}$ is 0 and out-degree of $v_{st}$ is 0 in $G_{sub}; $ iii) for each vertex $v_i \in V_{sub}$ with $v_i \neq v_{st}$, the successor vertices $V^s_i \subset V_{sub}$; iv) for each vertex $v_i \in V_{sub}$ with $v_i \neq v_{ss}$, the predecessor vertices $V^p_i \subset V_{sub}$.
\end{definition}

\begin{algorithm}
\footnotesize
\caption{Generate a One-Entry-One-Exit Subgraph} 
\label{alg:split_oeoe} 
\lstset{basicstyle=\ttfamily\footnotesize, breaklines=true}
\begin{tabular}{|lp{2.7in}|}\hline
\textbf{Input}:
    & A transformed iCFG $G(V,E)$ without cycle edges \\
    & Entry and exit vertices $v_{ss}, v_{st}$ of subgraph \\
\textbf{Output}:
    & A subgraph $G_{sub}(V_{sub},E_{sub})$ \\
    & The transformed graph $G'(V',E')$ from $G$ without $V_{sub}$ and $E_{sub}$ \\
\hline
\end{tabular}
\begin{algorithmic}[1]

    
    

\State $V_{sub} \gets \{v_{ss}, v_{st}\}$
\For{$v_k \in V, v_{ss} ~\mathrm{reaches}~ v_k, v_k ~\mathrm{reaches}~ v_{st}$}
    \State Add $v_k$ to $V_{sub}$
\EndFor

\State $G_{sub}(V_{sub},E_{sub}) \gets \mathrm{get\_induced\_graph(} G, V_{sub} \mathrm{)}$

\State $V'_{sub} \gets V_{sub} \backslash \{v_{ss}, v_{st}\}$
\State $V' \gets V\backslash V'_{sub}$
\State $E' \gets E\backslash E_{sub}$
\State Merge $v_{ss}$ and $v_{st}$ into $v_{sst}$ in $V'$    

\State
\Return $G_{sub}(V_{sub},E_{sub})$, $G'(V', E')$

\end{algorithmic}
\end{algorithm}

To ensure the provable correctness of the MPC computation in each one-entry-one-exit subgraph, we propose Theorem~\ref{thm:graph_split} and provide its proof in Appendix~\ref{app:graph_split}. After conceptually splicing the MPCs of the subgraphs at the merged vertex, the path cover for the transformed iCFG, excluding cycle edges, remains minimum. Regarding the maximum $k$ value, it can be determined through path cover enumeration, as described in Section~\ref{sec:m_mpc}. At run-time, a function call may appear multiple times along a single path. For each function call, \ToolName{} always considers a new path in the MPC of the corresponding function subgraph to guide path selection.

\begin{theorem} \label{thm:graph_split}
    For a directed acyclic graph $G(V,E)$, $G_{sub}(V_{sub},E_{sub})$ is a one-entry-one-exit subgraph of $G$. $G'(V', E')$ is a transformed graph from $G(V,E)$ computed by Algorithm \ref{alg:split_oeoe} and $v_{sst}$ is the merged virtual vertex. $P^{sub}_m$ is an MPC of $G_{sub}$; $P_m$ is an MPC of $G$. $P'_m$ is an MPC of $G'$ that satisfies: i) there are $k$ paths going through $v_{sst}$ in $P'_m$; ii) $k$ is the maximum among all MPCs in $G'$. We have $|P_m| = |P'_m|-k+\mathrm{max(}|P^{sub}_m|, k \mathrm{)}$.
\end{theorem}
See Appendix \ref{app:graph_split} for proof.

\vspace{0.1cm}
\noindent
\textbf{Transform Loop Cycles.} We adopt the definition of loops in LLVM~\cite{lattner2004llvm}, with some additional terminology illustrated in Figure~\ref{subfig:loop_trans_ori}~\cite{llvm0000loop}. According to this definition, a loop has a single header but may contain multiple exiting nodes within its body. Consequently, the one-entry-one-exit subgraph used for function calls cannot be applied to loops, necessitating the definition of a different type of subgraph. In addition to the common loops defined by LLVM, developers occasionally write extraordinary loops that have more than one header (entry) node. We do not handle these extraordinary loops for two reasons. First, extraordinary loops cannot be easily transformed into a subgraph suitable for computing MPCs and splicing at a merged vertex. Second, these loops are extremely rare in real-world programs, accounting for only 0.3\% of the benchmark programs, as shown in Section~\ref{sec:rq4}.

\revise{The second type of subgraphs introduced in this section is called a loop subgraph, as shown in Figure~\ref{fig:loop_trans}. Each loop contains at least one back edge. To represent the loop body (i.e., the cycle body), we remove the back edge and introduce a virtual representation as a merged vertex, as illustrated in Figure~\ref{subfig:loop_trans_vnode}. The transformed iCFG is then split into two parts, as shown in Figure~\ref{subfig:loop_trans_split}, where the right part represents a loop subgraph. A loop subgraph can be formally defined as a directed acyclic subgraph $G_{sub}(V_{sub}, E_{sub})$ in Definition~\ref{def:loop_subg}. The logic for generating a loop subgraph is described in Algorithm~\ref{alg:gen_loop_g}. Loop information, $loop\_info$, can be obtained using loop analysis provided by compilers such as LLVM.}

\begin{definition}[Loop Subgraph] \label{def:loop_subg}
    $G_{sub}(V_{sub}, E_{sub})$ is a loop subgraph of $G(V,E)$ iff. i) $G_{sub}$ is an induced subgraph of $G$; ii) $G_{sub}$ has a vertex $v_{ss}$ and a set of vertices $V_{st} = \{v_{st_1},...,v_{st_k}\}$ where in-degree of $v_{ss}$ is 0 and out-degree of $v_{st_i} \in V_{st}$ is 0 in $G_{sub}; $ iii) for each vertex $v_i \in V_{sub}$ with $v_i \notin V_{st}$, the successor vertices $V^s_i \subset V_{sub}$; iv) for each vertex $v_i \in V_{sub}$ with $v_i \neq v_{ss}$, the predecessor vertices $V^p_i \subset V_{sub}$.
\end{definition}


\begin{algorithm}
\footnotesize
\caption{Generate a Loop Subgraph} 
\label{alg:gen_loop_g} 
\lstset{basicstyle=\ttfamily\footnotesize, breaklines=true}
\begin{tabular}{|lp{2.7in}|}\hline
\textbf{Input}:
    & A transformed iCFG $G(V,E)$ without cycle edges \\
    & Loop info $loop\_info$ analyzed by the compiler \\
\textbf{Output}:
    & A loop subgraph $G_{sub}(V_{sub},E_{sub})$ \\
    & A transformed graph $G'(V',E')$ from $G$ without $V_{sub}$ and $E_{sub}$ \\
\hline
\end{tabular}
\begin{algorithmic}[1]

\State $v_{ss} \gets \mathrm{get\_header(}loop\_info\mathrm{)}$
\State $V_{st} \gets \mathrm{get\_exit\_vertices(}loop\_info\mathrm{)}$
\State $E_{st} \gets \mathrm{get\_exiting\_edges(}loop\_info\mathrm{)}$
\State $V_{sub} \gets \mathrm{get\_vertices(}loop\_info\mathrm{)}$
\State $E_{sub} \gets \mathrm{get\_edges(}loop\_info\mathrm{)}$

\State Replace $v_{ss}$ with a virtual vertex $v_{sst}$ in $G$
\State $V' \gets V\backslash V_{sub}$
\State $E' \gets E\backslash E_{sub}$

\State $V_{sub} \gets V_{sub} \cup V_{st}$ \AlgComment{Include virtual exit vertices}
\State $E_{sub} \gets E_{sub} \cup E_{st}$ \AlgComment{Include virtual exiting edges}

\For{$v_{st_i} \in V_{st}$}
    \State Add $(v_{sst},v_{st_i})$ to $E'$
\EndFor

\State
\Return $G_{sub}(V_{sub},E_{sub})$, $G'(V', E')$

\end{algorithmic}
\end{algorithm}

We present Theorem~\ref{thm:loop_graph_split} and provide its proof in Appendix~\ref{app:loop_graph_split} to establish the correctness of the computation within each loop subgraph. Similar to function calls, at run-time, a loop may appear multiple times along a single path. For each loop, \ToolName{} steers path selection toward the next cycle of the loop subgraph rather than breaking out of the loop. Within each loop, \ToolName{} considers a new path in the MPC of the loop subgraph to guide path selection until no matching path remains in the MPC. At that point, \ToolName{} selects a state that breaks out of the loop.

\begin{theorem} \label{thm:loop_graph_split}
    For a directed acyclic graph $G(V,E)$, $G_{sub}(V_{sub},E_{sub})$ is a loop subgraph of $G$. $G'(V',E')$ is a transformed graph from $(V,E)$ computed by Algorithm \ref{alg:gen_loop_g}; $v_{sst}$ is the replaced virtual vertex; $V_{st}$ is a group of exit vertices. $P^{sub}_m$ is an MPC of $G_{sub}$; $P_m$ is an MPC of $G$. $P'_m$ is an MPC of $G'$ that satisfies: i) $P'_{sst} \subset P'_m$ is a group of paths in which each path $p_{sst_i}$ goes through edge $(v_{sst},v_{st_i})$ with $v_{st_i} \in V_{st}$; ii) $k=|P'_{sst}|$ is the maximum among all MPCs in $G'$. We have $|P_m|=|P'_m|-k+\mathrm{max(} |P^{sub}_m, k| \mathrm{)}$.
\end{theorem}
See proof in Appendix \ref{app:loop_graph_split}.


\subsubsection{Multiple MPCs to Cover Diverse Paths} \label{sec:m_mpc}

\revise{We now extend the computation of a single MPC to the computation of multiple MPCs. Conceptually, an MPC in a directed graph is not unique, meaning that a graph may have multiple distinct MPCs. Algorithm~\ref{alg:sin_mpc_comp} outlines an approach to compute a single MPC, which is derived from a maximum matching. The generation of this matching depends on the starting vertex used in the maximum matching algorithm, making the computation of an MPC inherently non-deterministic among all possible MPCs in the graph. However, run-time path selection is influenced by various factors, including data flow and path feasibility determined by SMT solvers. As a result, it is not possible to predict which MPC aligns most closely with the actual path selection in the symbolic execution engine.}

To handle scenarios where paths are fixed and uniquely determined without alternatives, enumerating all MPCs is an effective approach, as it ensures that every possible path cover is considered. Algorithm~\ref{alg:sin_mpc_comp} employs a maximum matching method to compute an MPC using the Hopcroft-Karp algorithm~\cite{hopcroft1973n}. However, the maximum matching generated by the Hopcroft-Karp algorithm is non-deterministic, as the algorithm begins its iterations from a random vertex. Consequently, the MPC generated by Algorithm~\ref{alg:sin_mpc_comp} is also non-deterministic. Theorem~\ref{thm:mpc_matching} establishes that for each MPC in $G$, there exists a corresponding maximum matching in $G_b$. Therefore, the problem of enumerating MPCs in $G$ can be reduced to the problem of enumerating maximum matchings in $G_b$.

\begin{theorem} \label{thm:mpc_matching}
    For a directed acyclic graph $G(V,E)$, an MPC $P_m$ of $G$ and a transformed bipartite graph $G_b(V_b,E_b)$ of $G$ introduced in Algorithm \ref{alg:sin_mpc_comp}, there must be a maximum matching $M_m$ in $G_b$ that can be converted to $P_m$ via Algorithm \ref{alg:sin_mpc_comp}.
\end{theorem}
See Appendix \ref{app:mpc_matching} for proof.

Enumerating maximum matchings in a bipartite graph is a well-studied problem in graph theory~\cite{tsukiyama1977new, uno1997algorithms}. We adopt Takeaki's method~\cite{uno1997algorithms} to enumerate all maximum matchings in a bipartite graph. This method leverages the property that the symmetric difference between two maximum matchings consists of cycles and paths of even length. It generates all maximum matchings by exchanging edges in existing matchings and iteratively applying this process. The time complexity of this algorithm is $O(|E_b||V_b|^{\frac{1}{2}} + |V_b||max\_matching\_group|)$~\cite{uno1997algorithms}. To optimize this computation process, we reduce the time complexity through two approaches. First, we leverage the one-entry-one-exit subgraphs introduced in Definition~\ref{def:oeoe_subg} and Theorem~\ref{thm:graph_split}. By transforming a large, complex graph into smaller subgraphs, where each subgraph is a one-entry-one-exit graph, we enable independent analysis of each subgraph using Takeaki's method and Algorithm~\ref{alg:sin_mpc_comp} to generate multiple MPCs. Second, for highly complex subgraphs, we impose a limit on the number of MPCs generated to achieve an approximation. In practice, most subgraphs are small, containing only a few vertices and edges, which makes this optimization highly effective.





\subsubsection{Path Prioritization at Run-time} \label{sec:path_prior}

\revise{After transforming the iCFG and computing MPCs for each subgraph, we do not merge these MPCs into a single set for the entire iCFG, as this operation is computationally expensive. Instead, we consider the MPCs in each subgraph independently during the path prioritization phase at run-time.}

\revise{At run-time of symbolic execution, \ToolName{} maintains a group of MPCs for each subgraph. This group initially contains all enumerated MPCs generated during the preprocessing stage. The symbolic execution engine forks an execution state into two (or more) states at each branch. Each state represents a subpath from the program entry to the current basic block. After SMT solvers verify the feasibility of the two states, \ToolName{} compares the subpaths of the states with all paths in the MPCs in the group for their respective subgraphs. In practice, \ToolName{} only needs to compare subpaths in the subgraphs described in Section~\ref{sec:mpc_construct} and Section~\ref{sec:m_mpc}. \ToolName{} then selects the state with the matched subpath to continue execution. Meanwhile, \ToolName{} removes those MPCs that do not contain a matching path for the current subpath, but ensures that at least one MPC remains in the group. If none of the MPCs provides a matching path, \ToolName{} handles infeasible paths as described in Section~\ref{sec:inf_path}. Thus, on the one hand, the multiple MPCs provide a small subset of paths and prioritize these paths to guide path selection at run-time; on the other hand, run-time path feasibility influences the group of MPCs, dynamically refining the guidance provided by the MPCs.}

\subsection{Handle Infeasible Paths} \label{sec:inf_path}


In this section, we introduce another component of \ToolName{} designed to handle infeasible paths. The MPCs in \ToolName{} provide predefined paths to guide run-time path selection. However, despite enumerating MPCs, these paths are derived solely from iCFG and do not incorporate data-flow analysis. As a result, there is no guarantee that every run-time path will have a corresponding path in the MPC group. This is because the prefix of a path in the MPC group may become infeasible according to SMT solvers at run-time.

We approach the infeasible path problem from the perspective of conditional branches. A conditional branch contains a condition, which is a logical expression such as $x > y$. By altering the results of the branch condition, we can modify the constraints of a subpath, potentially changing its feasibility. The outcome of a conditional branch depends on the values assigned to its associated variables. These variables must have been defined in preceding basic blocks, which means that they depend on prior basic blocks and branches. Therefore, the key to identifying a path prefix that potentially reaches this branch lies in analyzing the dependence information of these variables.


\begin{definition}[Data Dependence] \label{def:data_dep}
    Suppose that a variable $var_i$ is used in the condition $c_i$ of a branch $br_i$. Data dependence is a map from $br_i$ to a basic block node $n_j$ in CFG. We say that $br_i$ has data dependence on $n_j$ iff. $n_j$ defines $var_i$ such that i) there is a path $p$ from $n_j$ to $br_i$, and ii) there is no any other node $n_k$ on $p$ that defines $var_i$.
\end{definition}

\begin{definition}[Potential Dependence] \label{def:pot_dep}
    Suppose a variable $var_i$ is used in the condition $c_i$ of a branch $br_i$. The potential dependence is a map from $br_i$ to a branch $br_j$. $br_i$ is potentially dependent on $br_j$ iff. i) there is a path $p$ from $br_j$ to $br_i$ in which $var_i$ is not defined; ii) there is another path $p'$ from $br_j$ to $br_i$ where $var_i$ is defined.
\end{definition}

Definition~\ref{def:data_dep} and Definition~\ref{def:pot_dep} introduce two classic types of program dependence that have been studied for decades~\cite{agrawal1993incremental, gyimothy1999efficient, qi2013path, wang2016dependence}. \textbf{Data dependence}, derived from the def-use chain in program analysis, represents the relationship between a branch condition and a basic block that may influence this condition, thereby affecting branch choices. \textbf{Potential dependence} reveals that a previous branch can influence the definition of a variable, which, in turn, can affect the outcome of the current branch. We leverage data dependence and potential dependence to identify the preceding basic blocks and branches that influence a given branch. We trace the operand variables involved in def-use chains and also track the parameters and returns of function calls. Additionally, we perform basic pointer analysis by ignoring nested or multilevel pointers, as complete pointer analysis is computationally expensive.

\revise{At run-time of symbolic execution, when \ToolName{} invokes the infeasible path handling mechanism, it identifies the last unvisited basic block on the infeasible path and its corresponding branch, denoted as \textit{br\_unvisited}. It then performs a backward search on the iCFG starting from \textit{br\_unvisited} until it discovers an ancestor basic block, \textit{bb\_ancestor}, that has a program dependence on \textit{br\_unvisited}. \ToolName{} locates symbolic execution states that reach \textit{bb\_ancestor} and redirects symbolic execution to these states. As described in Section~\ref{sec:path_prior}, \ToolName{} ignores execution states with mismatched subpaths in the MPC group during execution. However, during the current search for handling infeasible paths, \ToolName{} reconsiders these previously ignored states. If such a state is reconsidered, \ToolName{} temporarily disregards the current mismatched subpaths to allow the execution of this state to continue.}

\begin{table*}[!]
    \centering
    \caption{\textbf{\small The details of 12 benchmark programs. Each program is the latest version upon the evaluation. The number of basic blocks and code lines are calculated via KLEE internal coverage.}}
    \begin{tabular}[t]{llllrrr}
        \toprule
        \textbf{Project} & \textbf{Program} & \textbf{Version} & \textbf{Category} & \textbf{Binary Size} & \textbf{\# Basic Blocks} & \textbf{\# Code Lines} \\
        \midrule
         GNU bc & \ProgramStyle{bc} & 1.07.1 & Calculator & 169KB & 2430 & 3754 \\
         GNU ncurses & \ProgramStyle{tic} & 6.5 & Text & 606KB & 8972 & 9764 \\
         GNU make & \ProgramStyle{make} & 4.4.1 & Text & 562KB & 9450 & 11086 \\
         GNU bison & \ProgramStyle{bison} & 3.8.2 & Text & 1531KB & 22096 & 22690 \\
         GNU binutils & \ProgramStyle{readelf} & commit eb7892c4 & Binary & 2793KB & 38397 & 47939 \\
         GNU binutils & \ProgramStyle{strip-new} & commit eb7892c4 & Binary & 5417KB & 57257 & 76380 \\
         NASM & \ProgramStyle{nasm} & 2.16.02 & Binary & 2507KB & 16733 & 22369 \\
         libtiff & \ProgramStyle{tiffinfo} & 4.6.0 & Image & 887KB & 13490 & 16335 \\
         JasPer & \ProgramStyle{jasper} & 4.2.4 & Image & 1041KB & 14079 & 19146 \\
         Little CMS & \ProgramStyle{transicc} & 2.16 & Image & 929KB & 12242 & 17800 \\
         FLVMeta & \ProgramStyle{flvmeta} & 1.2.2 & Video & 361KB & 5122 & 6468 \\
         curl & \ProgramStyle{curl} & 8.10.1 & Network & 3458KB & 39413 & 46682 \\
        \bottomrule
    \end{tabular}
    \label{tab:bench}
\end{table*}

\section{Implementation}

\ToolName{} comprises two main components, all implemented in C++ using the LLVM\cite{lattner2004llvm} API (version 13.0.0). The first component is responsible for generating MPCs. We implement all algorithms mentioned in Section \ref{sec:mpc_construct} from scratch, including the Hopcroft-Karp algorithm\cite{hopcroft1973n}. The second component, for program dependence analysis, is built on the def-use chains provided by LLVM. Our implementation analyzes instructions one by one using a breadth-first search approach. However, we do not handle indirect calls due to the high cost associated with performing pointer analysis. Nonetheless, our dependence analysis remains efficient and operates on a small scale.

We now describe how \ToolName{} is integrated into one of the most widely used symbolic execution engines, KLEE~\cite{cadar2008klee} (version 3.1). We incorporate several C++ header and source files into KLEE's core modules. During the symbolic execution process, KLEE first loads a program's bitcode file and parses it into LLVM IR. \ToolName{} uses the LLVM IR to perform graph analysis and dependence analysis without modifying the IR. Once these analyses are complete, KLEE continues with the symbolic execution of the program. Path prioritization and infeasible path handling are embedded into KLEE's state update and state selection functions, as these functions serve as interfaces for new searchers. In Section~\ref{sec:overhead}, we evaluate the overhead introduced by \ToolName{} to the symbolic execution process.

\section{Evaluation}

We perform extensive evaluation of \ToolName{} aimed at answering the following research questions.
\begin{enumerate}
    \item \textbf{RQ1 (Code coverage and resource usage):} Can \ToolName{} improve code coverage and meanwhile reduce the number of paths and memory usage?
    \item \textbf{RQ2 (Finding security violations):} Can \ToolName{} find more security violations?
    \item \textbf{RQ3 (Runtime overhead):} What is the runtime overhead of \ToolName{}?
    \item \textbf{RQ4 (Design choice):} How do different components of \ToolName{} contribute to its performance?
\end{enumerate}

\subsection{Experiment Setup}

\begin{figure*}[!]
\centering
\captionsetup[subfloat]{captionskip=-.01cm, labelformat=empty}

\includegraphics[scale=0.6]{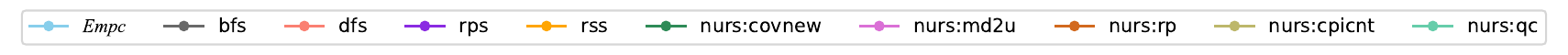}

\subfloat[\ProgramStyle{bc}]{
\includegraphics[width=0.23\textwidth]{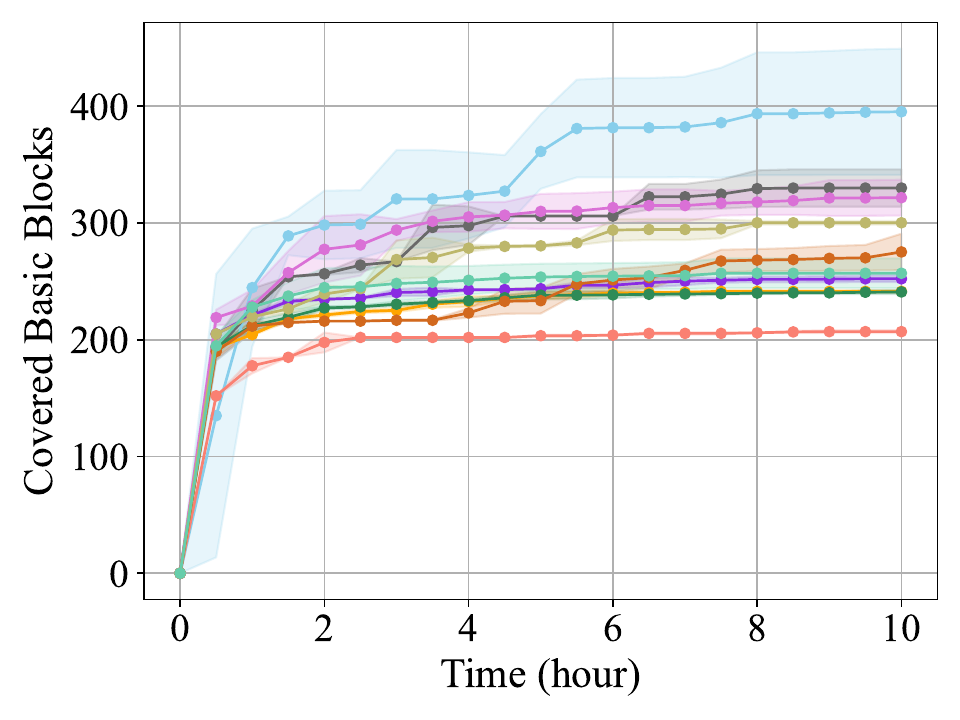}
\label{subfig:block_cov_bc}}
\hspace{-.1cm}
\subfloat[\ProgramStyle{tic}]{
\includegraphics[width=0.23\textwidth]{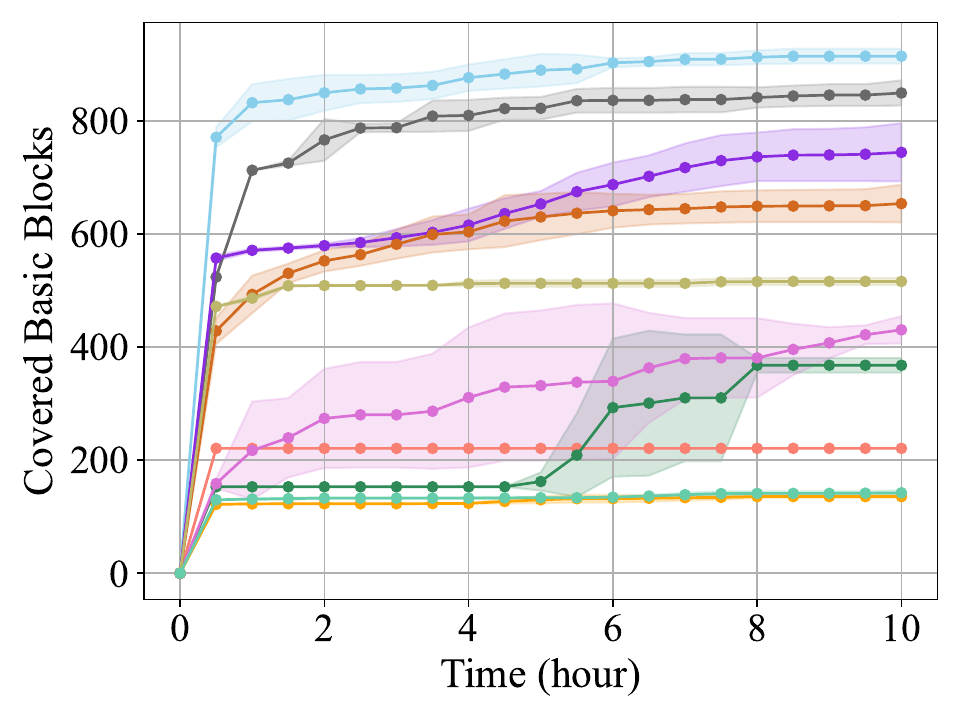}
\label{subfig:block_cov_tic}}
\hspace{-.1cm}
\subfloat[\ProgramStyle{make}]{
\includegraphics[width=0.23\textwidth]{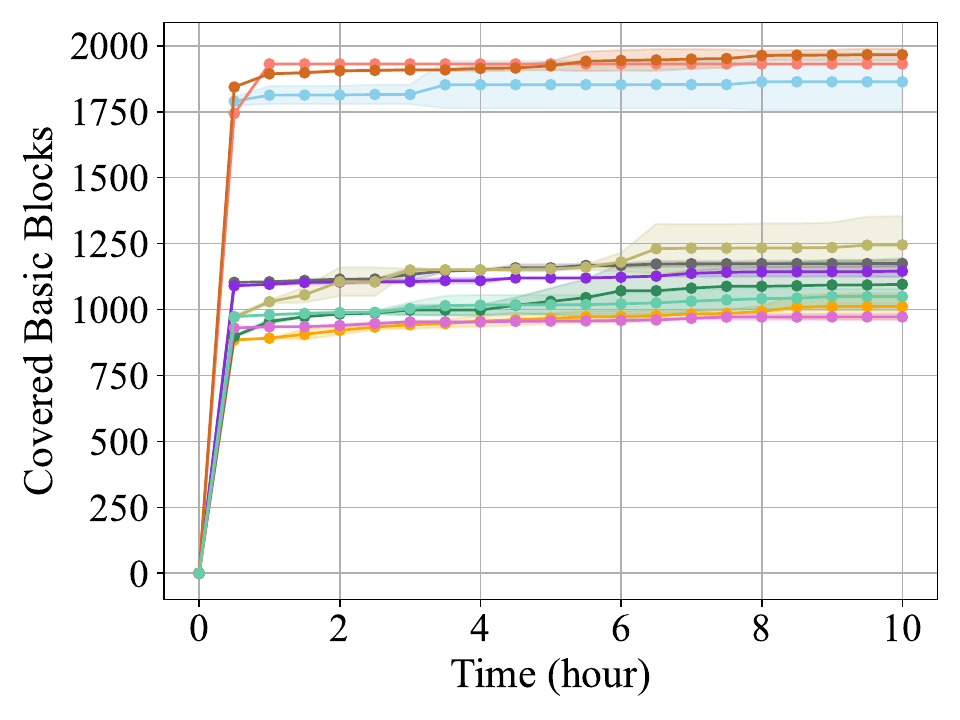}
\label{subfig:block_cov_make}}
\hspace{-.1cm}
\subfloat[\ProgramStyle{bison}]{
\includegraphics[width=0.23\textwidth]{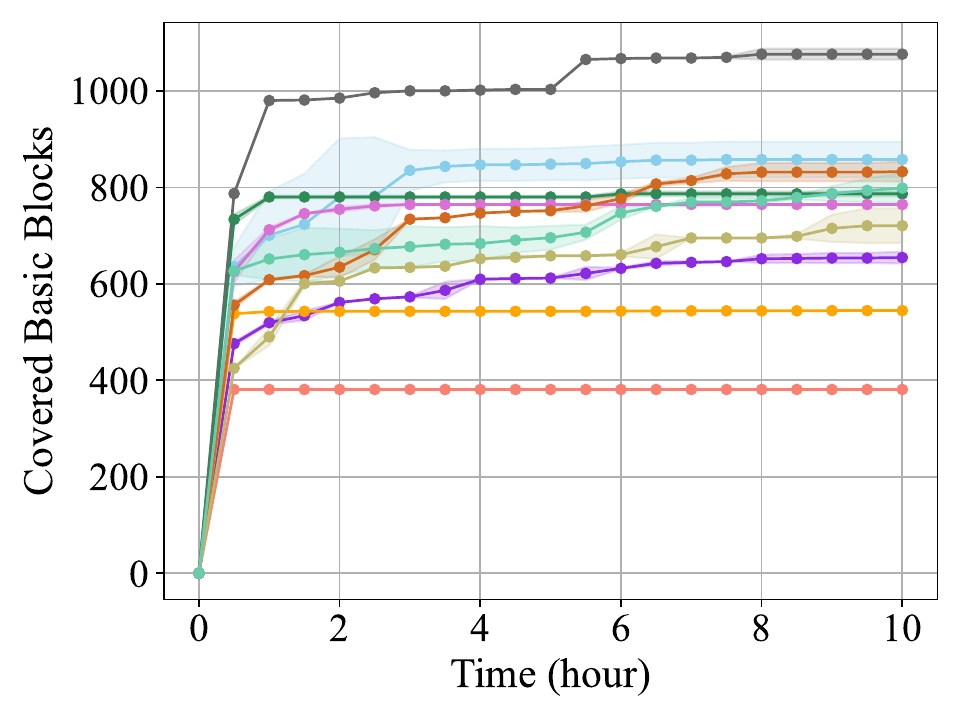}
\label{subfig:block_cov_bison}}

\subfloat[\ProgramStyle{readelf}]{
\includegraphics[width=0.23\textwidth]{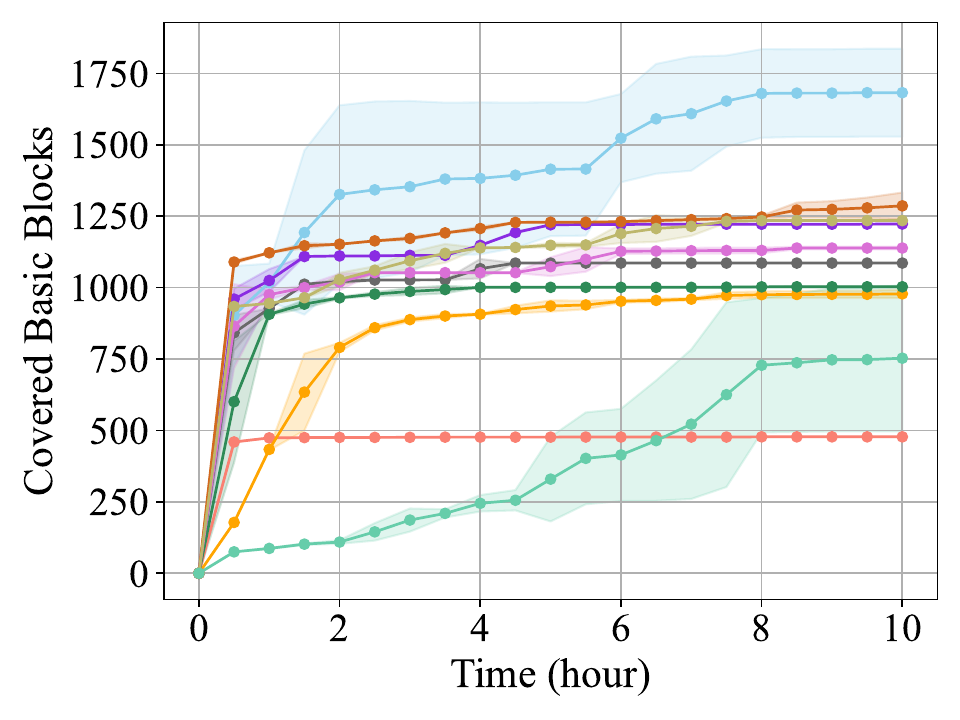}
\label{subfig:block_cov_readelf}}
\subfloat[\ProgramStyle{strip-new}]{
\includegraphics[width=0.23\textwidth]{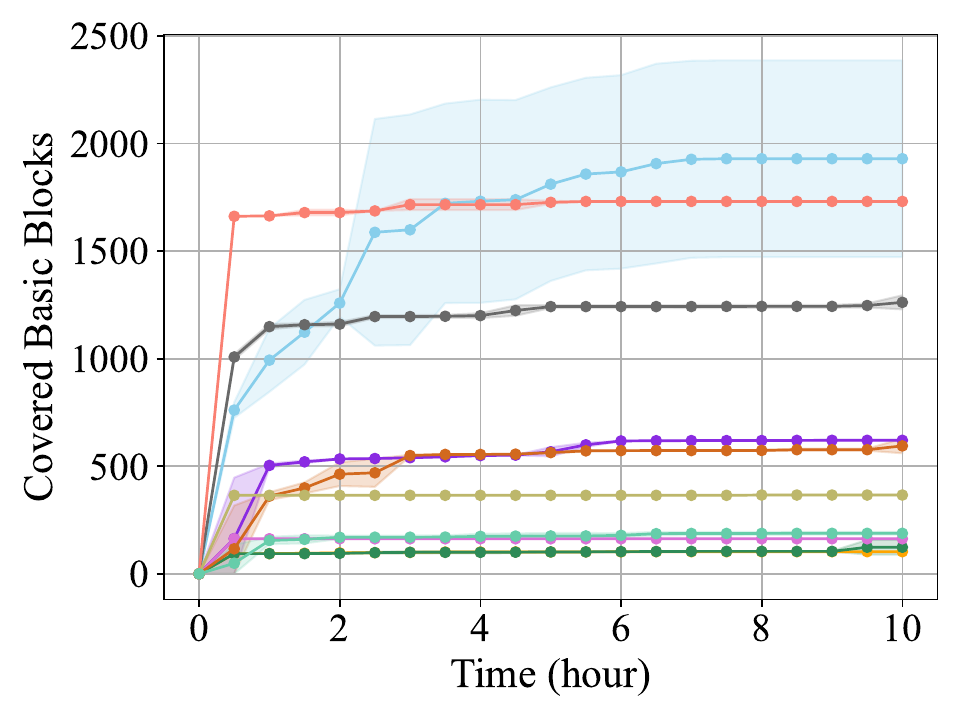}
\label{subfig:block_cov_strip-new}}
\hspace{-.1cm}
\subfloat[\ProgramStyle{nasm}]{
\includegraphics[width=0.23\textwidth]{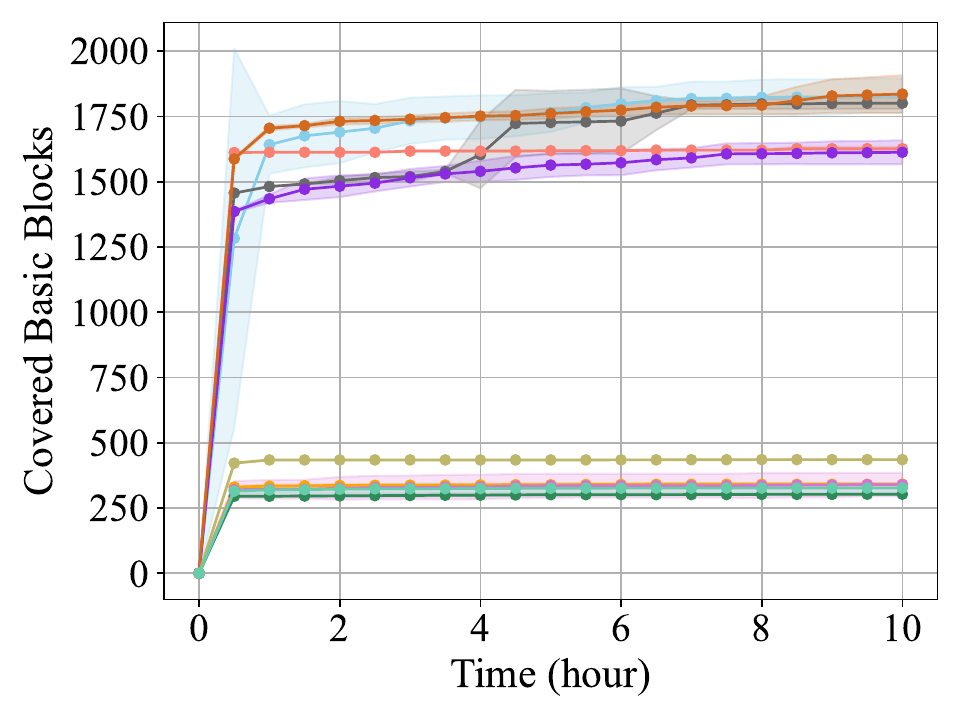}
\label{subfig:block_cov_nasm}}
\hspace{-.1cm}
\subfloat[\ProgramStyle{tiffinfo}]{
\includegraphics[width=0.23\textwidth]{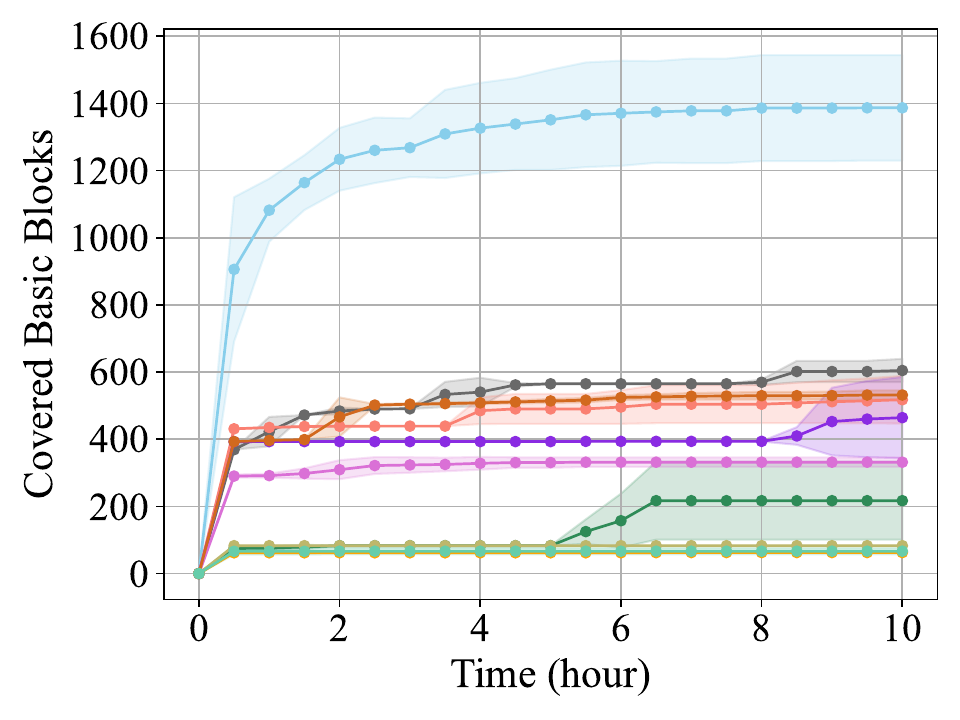}
\label{subfig:block_cov_tiffinfo}}

\subfloat[\ProgramStyle{jasper}]{
\includegraphics[width=0.23\textwidth]{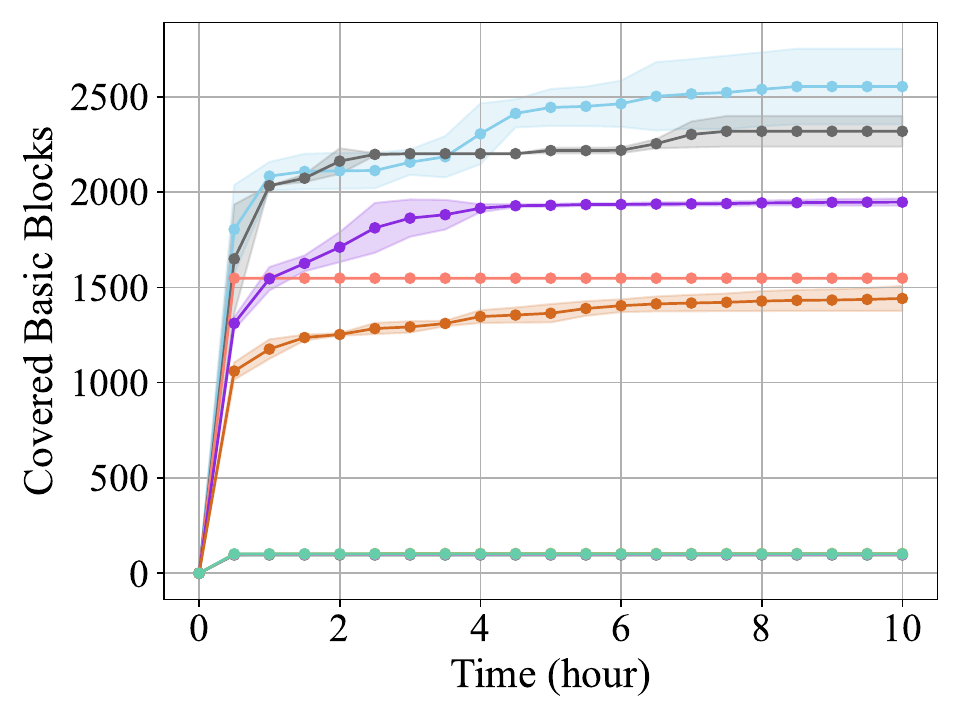}
\label{subfig:block_cov_jasper}}
\hspace{-.1cm}
\subfloat[\ProgramStyle{transicc}]{
\includegraphics[width=0.23\textwidth]{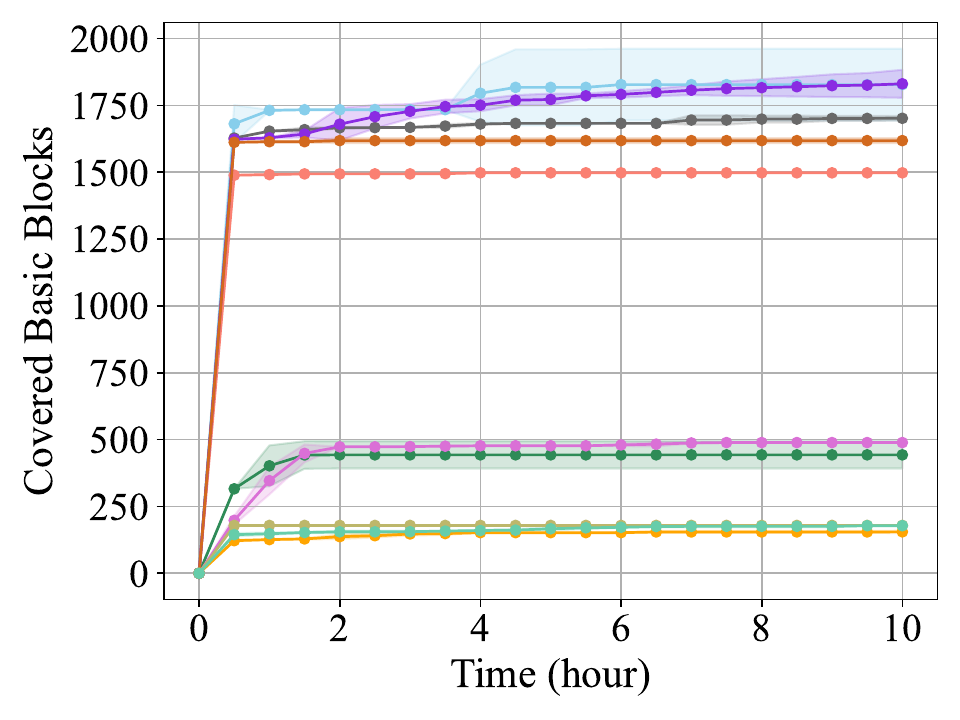}
\label{subfig:block_cov_transicc}}
\hspace{-.1cm}
\subfloat[\ProgramStyle{flvmeta}]{
\includegraphics[width=0.23\textwidth]{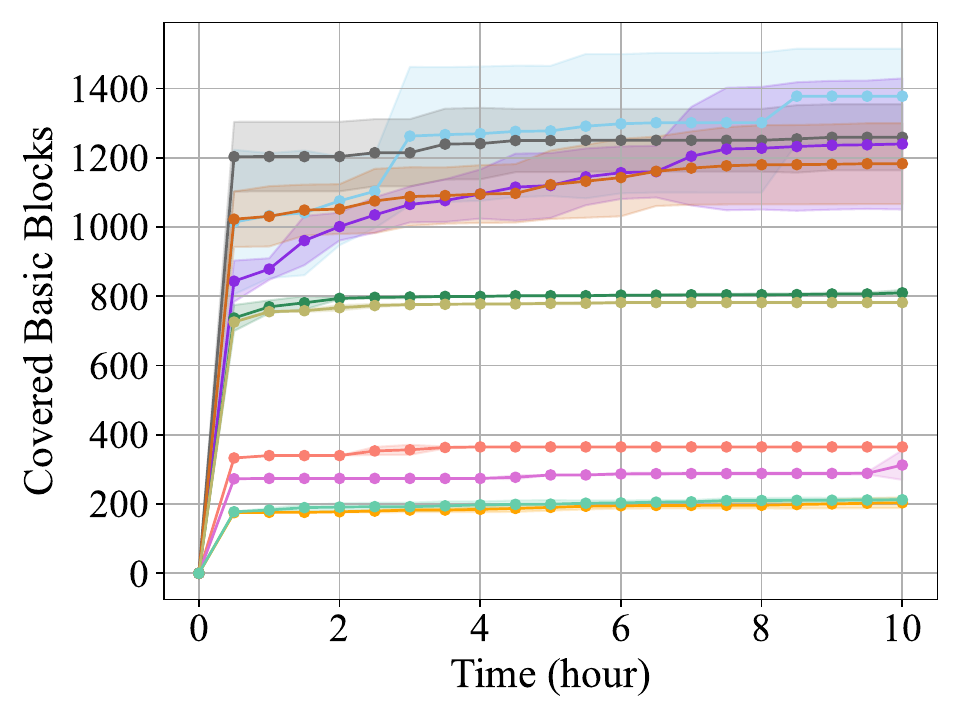}
\label{subfig:block_cov_flvmeta}}
\hspace{-.1cm}
\subfloat[\ProgramStyle{curl}]{
\includegraphics[width=0.23\textwidth]{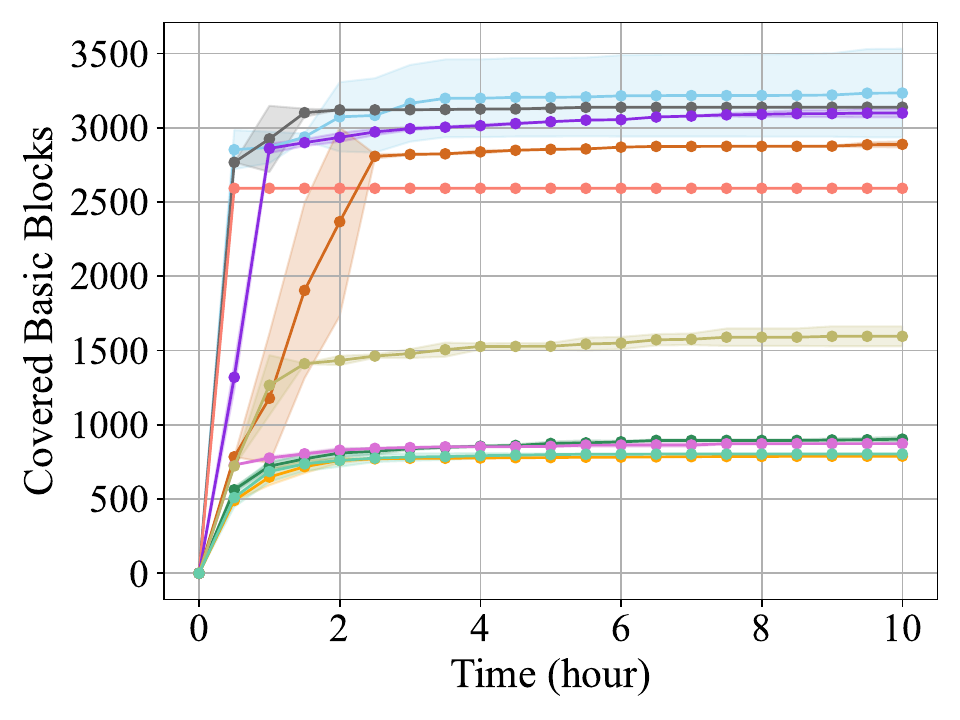}
\label{subfig:block_cov_curl}}

\caption{\textbf{\small The arithmetic mean internal coverage of basic blocks of \ToolName{} and all KLEE baseline search strategies running for 10 hours on 12 benchmark programs and one standard deviation error bars over 10 runs. The basic block coverage is computed via KLEE internal coverage.}}
\label{fig:block_cov}
\end{figure*}

\vspace{0.1cm}
\noindent
\textbf{Baseline search strategies.} We consider all 11 search strategies in KLEE~\cite{cadar2008klee, klee0000doc}, including \SearchToolStyle{bfs}, \SearchToolStyle{dfs}, \SearchToolStyle{random-state}, \SearchToolStyle{random-path}, \SearchToolStyle{nurs:rp}, \SearchToolStyle{nurs:depth}, \SearchToolStyle{nurs:covnew}, \SearchToolStyle{nurs:md2u}, \SearchToolStyle{nurs:icnt}, \SearchToolStyle{nurs:cpicnt}, and \SearchToolStyle{nurs:qc}. However, we exclude \SearchToolStyle{nurs:depth} and \SearchToolStyle{nurs:icnt} for the following reasons: \SearchToolStyle{nurs:depth} is similar to \SearchToolStyle{nurs:rp}, as both are guided by depth, but \SearchToolStyle{nurs:rp} consistently performs better than \SearchToolStyle{nurs:depth}. Similarly, \SearchToolStyle{nurs:icnt} is excluded for analogous reasons. In addition to the built-in search strategies in KLEE, we compare \ToolName{} with representative related works from recent years, as summarized in the following.
\begin{itemize}
    \item \SearchToolStyle{sgs}. Subpath-guided search (\SearchToolStyle{sgs})~\cite{li2013steering} prioritizes the selection of states whose subpaths have been explored the least frequently. In their implementation and evaluation, they executed four independent instances of the searcher with subpath lengths of 1, 2, 4, and 8. Each instance was allocated a quarter of the total time limit. We adopt their evaluation method as one of our baseline search strategies.
    \item \textsc{Learch}. \textsc{Learch}~\cite{he2021learning} is a novel learning-based strategy designed to effectively select promising states for symbolic execution, addressing the path explosion problem. It leverages existing heuristics for training data generation and feature extraction. \textsc{Learch} provides pre-trained feedforward network models~\cite{learch0000git}, which we use directly in our evaluation.
    \item \SearchToolStyle{cgs}. Concrete-constraint-guided search (\SearchToolStyle{cgs})~\cite{sun2024concrete} introduces a symbolic execution strategy guided by concrete constraints, aimed at covering more concrete branches and thereby improving overall code coverage in symbolic execution. \SearchToolStyle{cgs} leverages data dependence to prioritize states that are likely to traverse partially covered concrete branches. We adopt their approach in our evaluation.
\end{itemize}
The three methods described above address the path explosion problem from different perspectives. \SearchToolStyle{sgs} utilizes control-flow guidance and analyzes paths in CFGs. \textsc{Learch} represents the use of machine learning methods in symbolic execution. \SearchToolStyle{cgs} is a recent approach that focuses on heuristics derived from real-world programs and employs data-flow analysis in symbolic execution. These search strategies make our baselines more comprehensive and extensive. However, these strategies are implemented on older versions of KLEE and LLVM. For example, \textsc{Learch} is built on KLEE 2.1 and LLVM 6.0, while \SearchToolStyle{cgs} is based on KLEE 2.3 and LLVM 11.1.0. In contrast, \ToolName{} is implemented using KLEE 3.1 and LLVM 13.0. To address the issue of mismatched tool versions, we directly ported the source code of \SearchToolStyle{sgs} into our KLEE engine, as its implementation is relatively simple and the codebase is small. On the other hand, \textsc{Learch} and \SearchToolStyle{cgs} involve extensive instrumentation of KLEE's source code, making it difficult to port them directly. Fortunately, both provide source code and pre-configured environments that we can utilize. As a result, we build and maintain three separate environments separately for \ToolName{}, \textsc{Learch} and \SearchToolStyle{cgs}.


\vspace{0.1cm}
\noindent
\textbf{Benchmark programs.} We use 12 real-world open-source programs to evaluate \ToolName{}. These programs are widely used in fuzzing and symbolic execution techniques~\cite{fuzzbench,unifuzz-li,bohme2017directed,busse2020running,he2021learning,kapus2020pending,sun2024concrete}, and are applied in various domains such as text editing, binary operations, image and video processing, and networks. Table~\ref{tab:bench} provides details about these benchmark programs. To ensure \ToolName{} is applicable to current real-world programs, we use the latest versions of these benchmarks. Since \textsc{Learch} and \SearchToolStyle{cgs} must be evaluated in their respective environments, we ensure that the benchmark programs are configured and built with consistent parameters across all environments. KLEE provides symbolic environment settings for each executed program, where it symbolizes program arguments as well as certain input and output files based on these symbolic environment configurations. We configure these symbolic environments for our benchmarks as detailed in Table~\ref{tab:bench_klee_sym} in Appendix~\ref{app:bench_args}. We define symbolic arguments based on prior works~\cite{he2021learning,kapus2020pending,sun2024concrete} and usage information provided by the programs themselves.



\begin{figure*}[!]
\centering
\captionsetup[subfloat]{captionskip=-.01cm, labelformat=empty}

\includegraphics[scale=0.6]{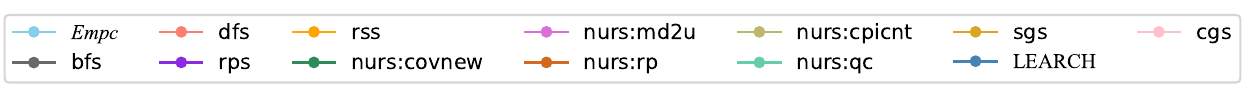}

\subfloat[\ProgramStyle{bc}]{
\includegraphics[width=0.23\textwidth]{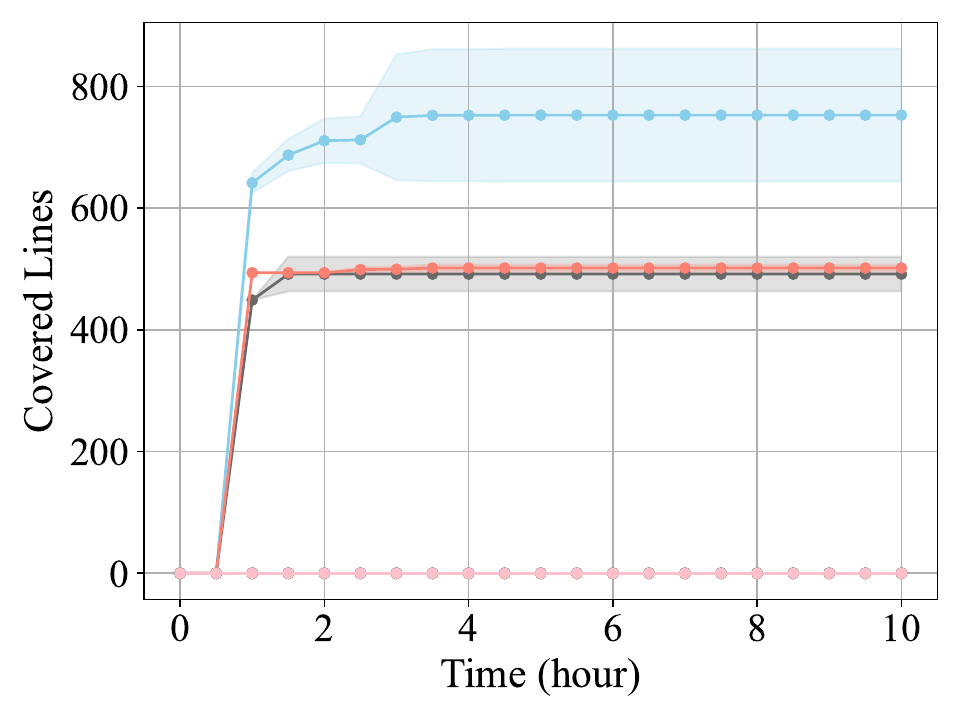}
\label{subfig:line_cov_bc}}
\hspace{-.1cm}
\subfloat[\ProgramStyle{tic}]{
\includegraphics[width=0.23\textwidth]{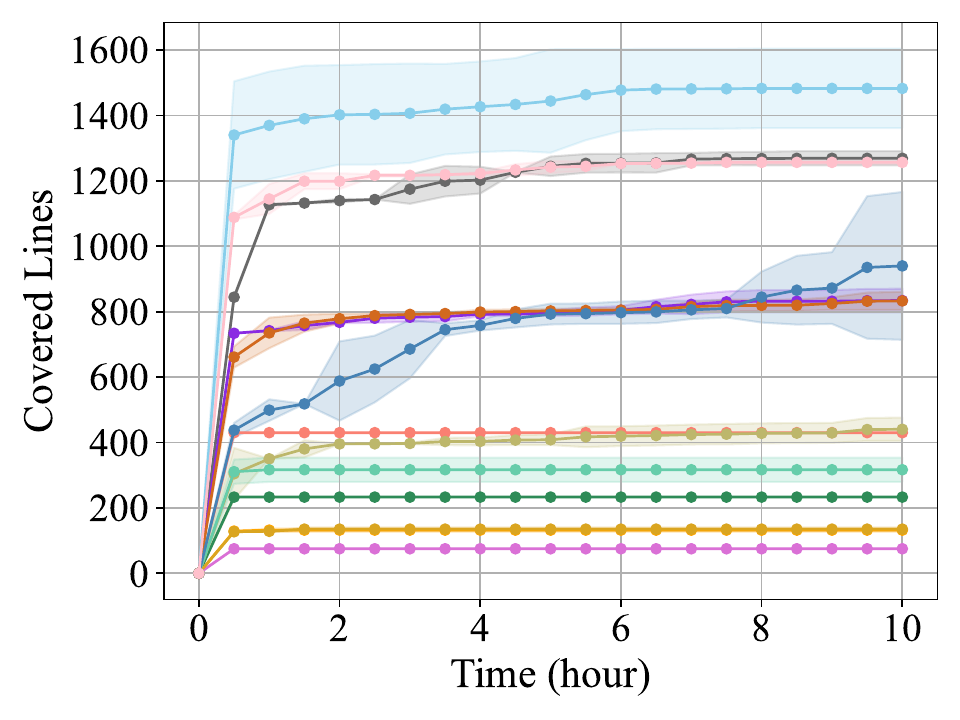}
\label{subfig:line_cov_tic}}
\hspace{-.1cm}
\subfloat[\ProgramStyle{make}]{
\includegraphics[width=0.23\textwidth]{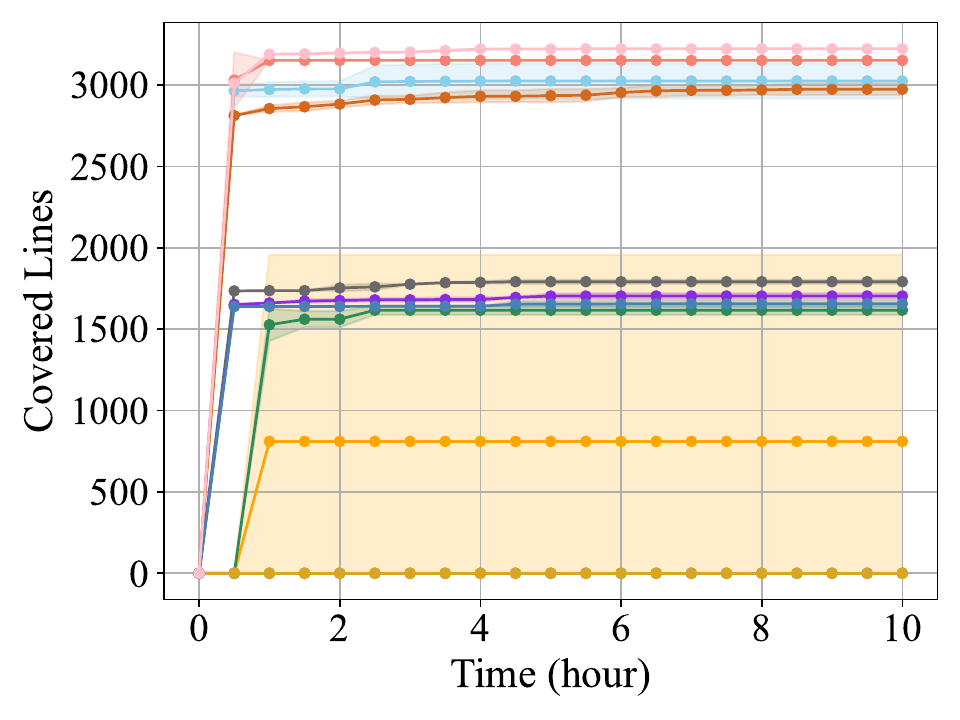}
\label{subfig:line_cov_make}}
\hspace{-.1cm}
\subfloat[\ProgramStyle{bison}]{
\includegraphics[width=0.23\textwidth]{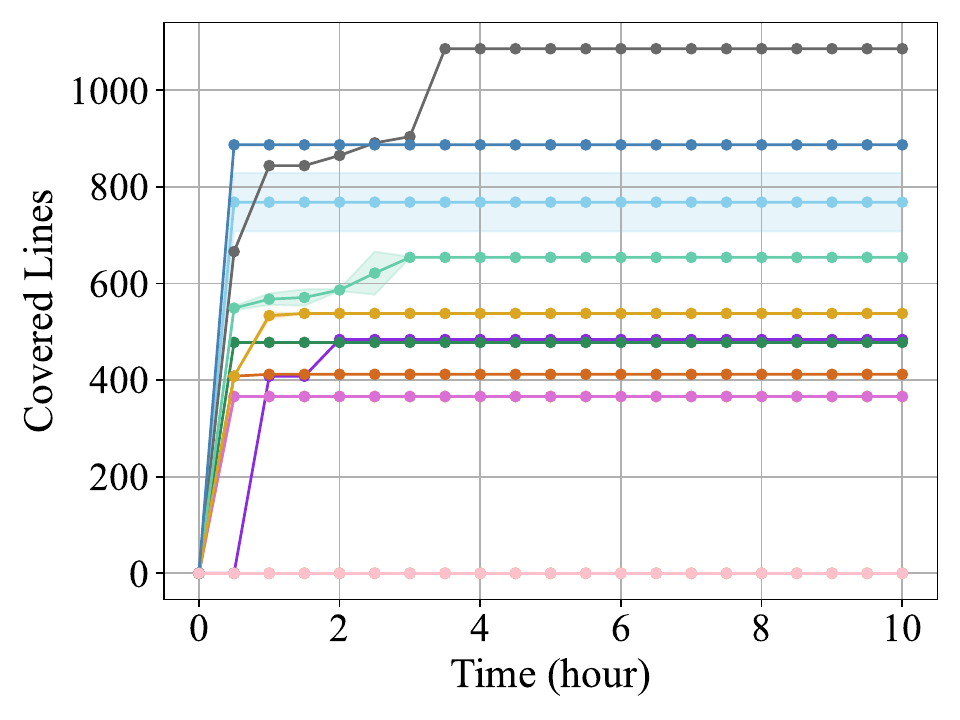}
\label{subfig:line_cov_bison}}

\subfloat[\ProgramStyle{tiffinfo}]{
\includegraphics[width=0.23\textwidth]{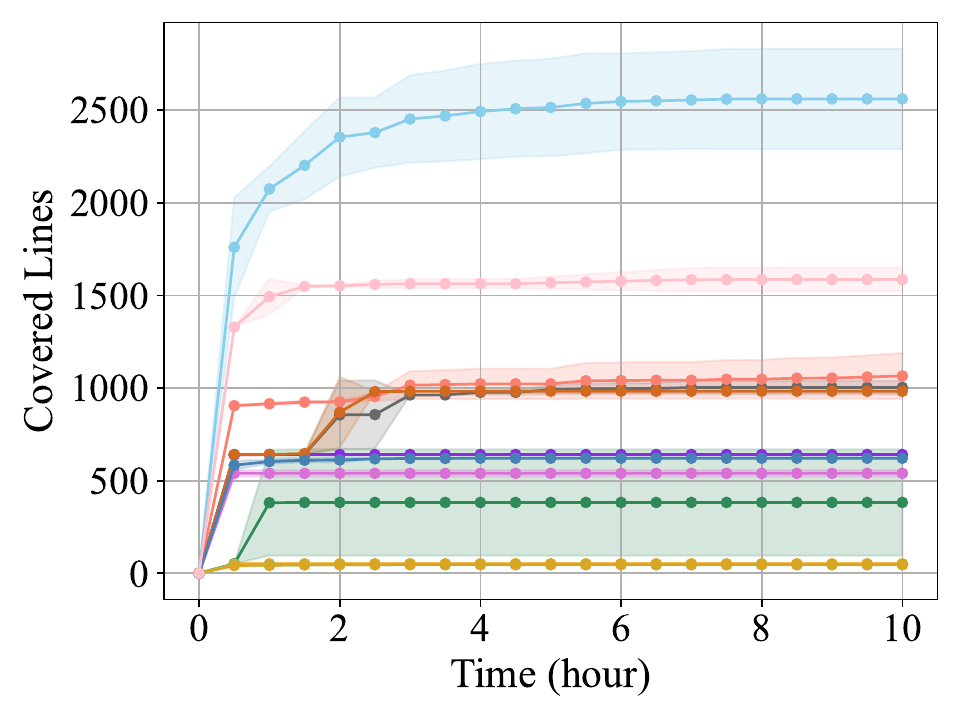}
\label{subfig:line_cov_tiffinfo}}
\hspace{-.1cm}
\subfloat[\ProgramStyle{jasper}]{
\includegraphics[width=0.23\textwidth]{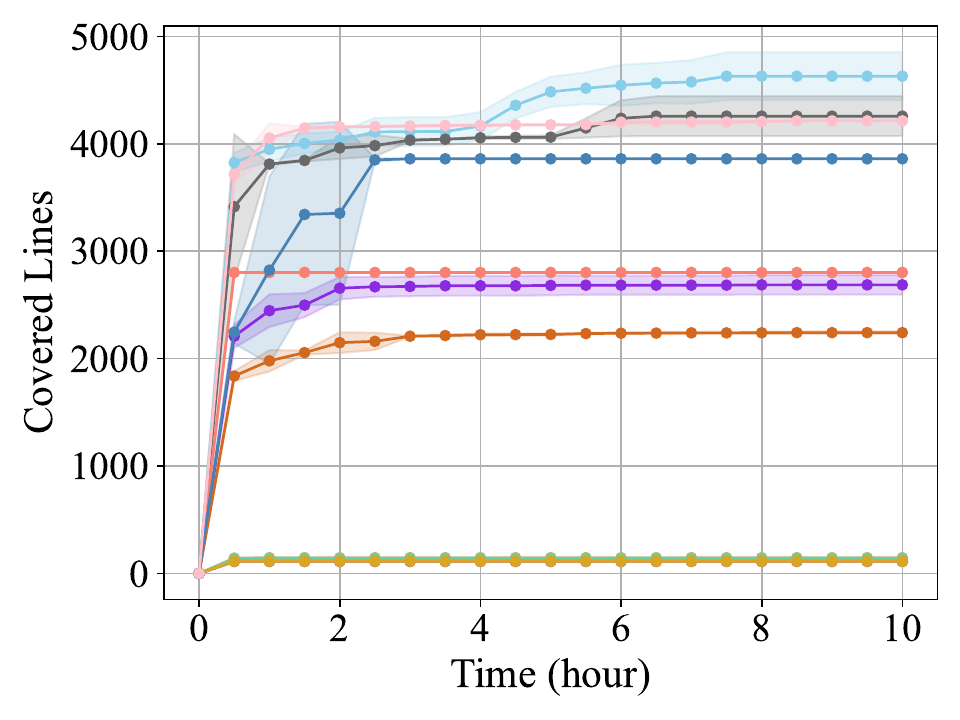}
\label{subfig:line_cov_jasper}}
\hspace{-.1cm}
\subfloat[\ProgramStyle{transicc}]{
\includegraphics[width=0.23\textwidth]{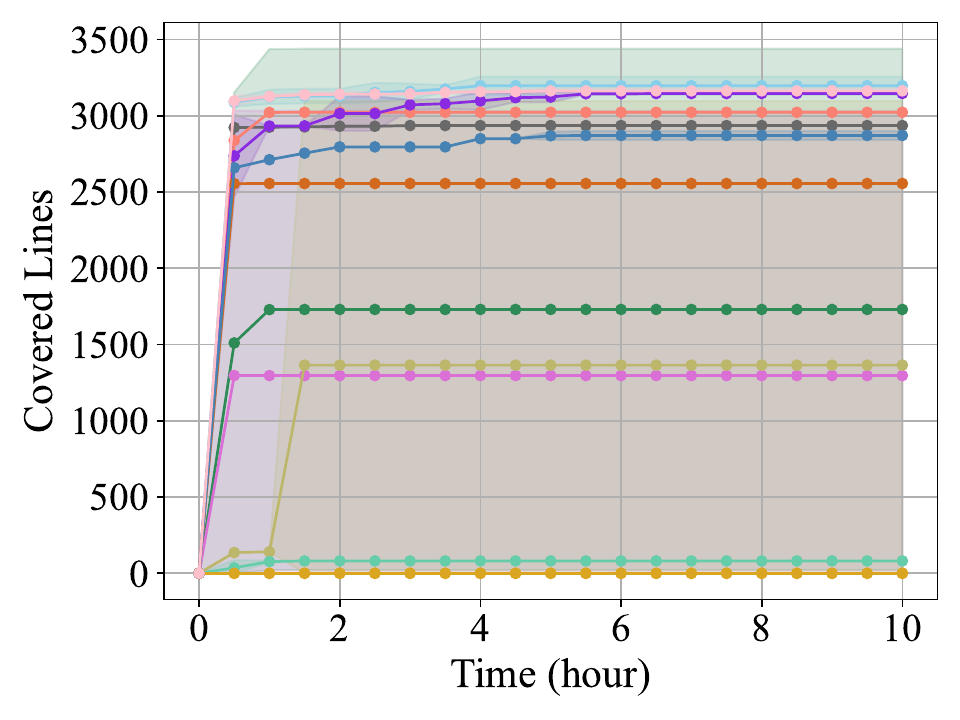}
\label{subfig:line_cov_transicc}}
\hspace{-.1cm}
\subfloat[\ProgramStyle{flvmeta}]{
\includegraphics[width=0.23\textwidth]{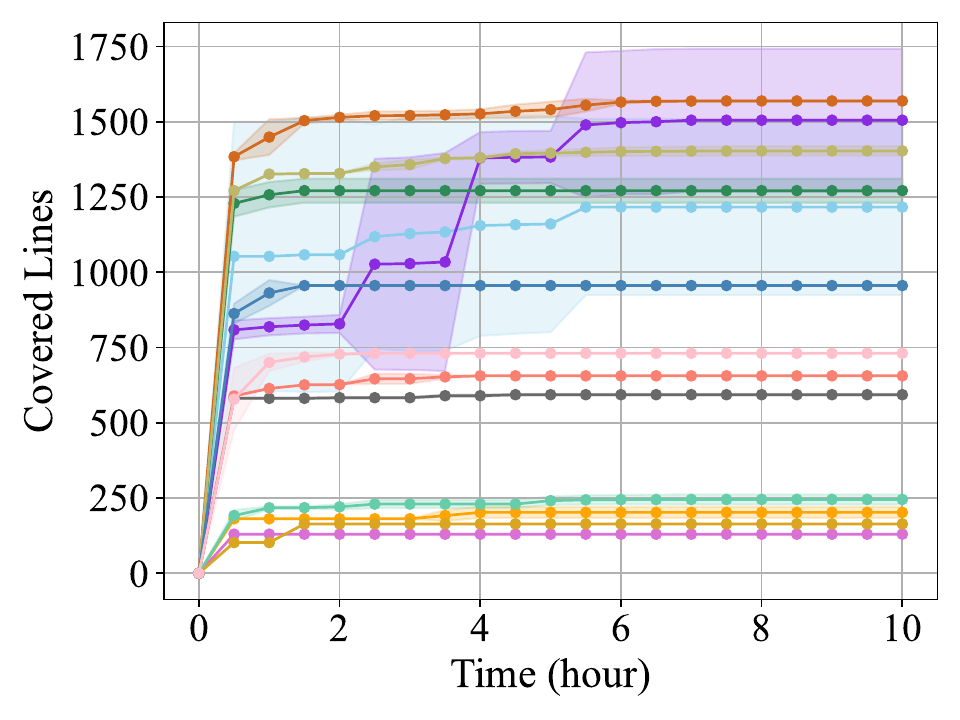}
\label{subfig:line_cov_flvmeta}}

\caption{\textbf{\small The arithmetic mean external coverage of code lines of \ToolName{}, KLEE search strategies and 3 prior works running for 10 hours on 8 benchmark programs and one standard deviation error bars over 10 runs. The line coverage is computed via replaying test inputs using \ProgramStyle{gcov}.}}
\label{fig:line_cov}
\end{figure*}

\vspace{0.1cm}
\noindent
\textbf{Environment setup.} We conduct all evaluations on a 64-bit machine equipped with 128 Intel Xeon (Cascade Lake) Platinum 8269CY CPUs and 3072 GB of RAM. Each KLEE instance is restricted to a single CPU core and a maximum of 32 GB of RAM. We run up to 95 KLEE instances simultaneously.

\subsection{RQ1: Code Coverage and Resource Usage} \label{sec:code_cov}

In this section, we evaluate \ToolName{} by comparing its code coverage and resource usage against the baseline search strategies mentioned earlier. Each KLEE instance is assigned a memory limit of 32 GB. This limit, which is significantly higher than the default of 4 GB, allows us to measure the resource usage, including memory consumption, for all search strategies. \textsc{Learch} and \SearchToolStyle{cgs} are executed in their respective environments based on older versions of KLEE, where some programs in our benchmarks (using the latest versions) cannot run successfully. These programs include \ProgramStyle{readelf}, \ProgramStyle{strip-new}, \ProgramStyle{nasm}, and \ProgramStyle{curl}. To address this compatibility issue, we divide the benchmarks and baselines into two groups: \textbf{Group A:} All benchmark programs with KLEE's baseline search strategies; \textbf{Group B:} A subset of 8 benchmark programs compatible with all baseline search strategies. We repeat each experiment 10 times, with each KLEE instance running for 10 hours. Finally, we compute the mean and standard deviation to analyze the results.

\begin{table*}[!]
    \centering
    \caption{\textbf{\small The size of memory (in bytes) held in the heap by a KLEE instance is measured at the 3rd, 7th, and 10th hours of execution. This evaluation is conducted only on KLEE baseline search strategies. Reduction*: We exclude comparisons with \SearchToolSmallStyle{dfs} due to its specificity. We calculate the proportion of states reduced by \ToolName{} relative to the strategy with the minimum number of states. Non-negative reductions are represented by green cells. The corresponding memory usage of \ToolName{} is shown in blue cells, while the memory usage of the compared strategy is shown in red cells.}}
\begin{tabular}{|l|l|c|c|c|c|c|c|ccccc|}
\hline
\multicolumn{1}{|c|}{\multirow{2}{*}{Program}} & \multirow{2}{*}{Time} & \multirow{2}{*}{Reduction*} & \multirow{2}{*}{\ToolName{}} & \multirow{2}{*}{\SearchToolSmallStyle{bfs}} & \multirow{2}{*}{\SearchToolSmallStyle{dfs}*} & \multirow{2}{*}{\SearchToolSmallStyle{rss}} & \multirow{2}{*}{\SearchToolSmallStyle{rps}} & \multicolumn{5}{c|}{\SearchToolSmallStyle{nurs}} \\ \cline{9-13} 
\multicolumn{1}{|c|}{} &  &  &  &  &  &  &  & \multicolumn{1}{c|}{\SearchToolSmallStyle{rp}} & \multicolumn{1}{c|}{\SearchToolSmallStyle{covnew}} & \multicolumn{1}{c|}{\SearchToolSmallStyle{md2u}} & \multicolumn{1}{c|}{\SearchToolSmallStyle{cpicnt}} & \SearchToolSmallStyle{qc} \\ \hline
\multirow{3}{*}{bc} & 3h & -38.2\% & 94M & 89M & 71M & 74M & 80M & \multicolumn{1}{c|}{68M} & \multicolumn{1}{c|}{87M} & \multicolumn{1}{c|}{31.3G} & \multicolumn{1}{c|}{427M} & 19.2G \\ \cline{2-13} 
 & 7h & -17.9\% & 99M & 126M & 75M & 105M & 101M & \multicolumn{1}{c|}{84M} & \multicolumn{1}{c|}{164M} & \multicolumn{1}{c|}{31.3G} & \multicolumn{1}{c|}{1.36G} & 31.3G \\ \cline{2-13} 
 & 10h & -10.2\% & 108M & 184M & 79M & 147M & 113M & \multicolumn{1}{c|}{98M} & \multicolumn{1}{c|}{237M} & \multicolumn{1}{c|}{31.3G} & \multicolumn{1}{c|}{2.37G} & 31.3G \\ \hline
\multirow{3}{*}{tic} & 3h & \ReduCellColor{44.9\%} & \ToolCellColor{14.6G} & 31.3G & 107M & 31.3G & 31.3G & \multicolumn{1}{c|}{31.3G} & \multicolumn{1}{c|}{\CompCellColor{26.5G}} & \multicolumn{1}{c|}{31.3G} & \multicolumn{1}{c|}{31.3G} & 31.3G \\ \cline{2-13} 
 & 7h & \ReduCellColor{16.3\%} & \ToolCellColor{26.2G} & \CompCellColor{31.3G} & 108M & 31.3G & 31.3G & \multicolumn{1}{c|}{31.3G} & \multicolumn{1}{c|}{31.3G} & \multicolumn{1}{c|}{31.3G} & \multicolumn{1}{c|}{31.3G} & 31.3G \\ \cline{2-13} 
 & 10h & \ReduCellColor{0\%} & \ToolCellColor{31.3G} & \CompCellColor{31.3G} & 106M & 31.3G & 31.3G & \multicolumn{1}{c|}{31.3G} & \multicolumn{1}{c|}{31.3G} & \multicolumn{1}{c|}{31.3G} & \multicolumn{1}{c|}{31.3G} & 31.3G \\ \hline
\multirow{3}{*}{make} & 3h & \ReduCellColor{68.7\%} & \ToolCellColor{727M} & 31.3G & 143M & 31.3G & 31.3G & \multicolumn{1}{c|}{31.3G} & \multicolumn{1}{c|}{31.3G} & \multicolumn{1}{c|}{31.3G} & \multicolumn{1}{c|}{\CompCellColor{2.32G}} & 31.3G \\ \cline{2-13} 
 & 7h & \ReduCellColor{84.9\%} & \ToolCellColor{1.11G} & 31.3G & 146M & 31.3G & 31.3G & \multicolumn{1}{c|}{31.3G} & \multicolumn{1}{c|}{31.3G} & \multicolumn{1}{c|}{31.3G} & \multicolumn{1}{c|}{\CompCellColor{7.35G}} & 31.3G \\ \cline{2-13} 
 & 10h & \ReduCellColor{87.6\%} & \ToolCellColor{1.19G} & 31.3G & 148M & 31.3G & 31.3G & \multicolumn{1}{c|}{31.3G} & \multicolumn{1}{c|}{31.3G} & \multicolumn{1}{c|}{31.3G} & \multicolumn{1}{c|}{\CompCellColor{9.61G}} & 31.3G \\ \hline
\multirow{3}{*}{bison} & 3h & -68.7\% & 361M & 274M & 188M & 282M & 246M & \multicolumn{1}{c|}{240M} & \multicolumn{1}{c|}{327M} & \multicolumn{1}{c|}{328M} & \multicolumn{1}{c|}{214M} & 302M \\ \cline{2-13} 
 & 7h & -61.5\% & 378M & 354M & 190M & 349M & 283M & \multicolumn{1}{c|}{278M} & \multicolumn{1}{c|}{396M} & \multicolumn{1}{c|}{387M} & \multicolumn{1}{c|}{234M} & 449M \\ \cline{2-13} 
 & 10h & -51.2\% & 381M & 385M & 192M & 400M & 301M & \multicolumn{1}{c|}{305M} & \multicolumn{1}{c|}{435M} & \multicolumn{1}{c|}{437M} & \multicolumn{1}{c|}{252M} & 529M \\ \hline
\multirow{3}{*}{readelf} & 3h & -26.4\% & 1.77G & 1.40G & 420M & 3.52G & 3.09G & \multicolumn{1}{c|}{2.11G} & \multicolumn{1}{c|}{3.77G} & \multicolumn{1}{c|}{2.38G} & \multicolumn{1}{c|}{1.75G} & 9.02G \\ \cline{2-13} 
 & 7h & -127\% & 4.11G & 1.81G & 604M & 5.43G & 4.56G & \multicolumn{1}{c|}{2.94G} & \multicolumn{1}{c|}{7.38G} & \multicolumn{1}{c|}{3.01G} & \multicolumn{1}{c|}{3.43G} & 31.3G \\ \cline{2-13} 
 & 10h & -204\% & 6.18G & 2.03G & 718M & 8.54G & 4.57 & \multicolumn{1}{c|}{4.27G} & \multicolumn{1}{c|}{10.0G} & \multicolumn{1}{c|}{3.60G} & \multicolumn{1}{c|}{4.35G} & 31.3G \\ \hline
\multirow{3}{*}{strip-new} & 3h & -203\% & 1.32G & 651M & 744M & 31.3G & 436M & \multicolumn{1}{c|}{461M} & \multicolumn{1}{c|}{31.3G} & \multicolumn{1}{c|}{31.3G} & \multicolumn{1}{c|}{31.3G} & 461M \\ \cline{2-13} 
 & 7h & -210\% & 1.43G & 841M & 750M & 31.3G & 461M & \multicolumn{1}{c|}{525M} & \multicolumn{1}{c|}{31.3G} & \multicolumn{1}{c|}{31.3G} & \multicolumn{1}{c|}{31.3G} & 470M \\ \cline{2-13} 
 & 10h & -201\% & 1.43G & 1.01G & 761M & 31.3G & 475M & \multicolumn{1}{c|}{571M} & \multicolumn{1}{c|}{31.3G} & \multicolumn{1}{c|}{31.3G} & \multicolumn{1}{c|}{31.3G} & 476M \\ \hline
\multirow{3}{*}{nasm} & 3h & \ReduCellColor{72.8\%} & \ToolCellColor{1.95G} & 31.3G & 644M & 11.8G & 31.3G & \multicolumn{1}{c|}{31.3G} & \multicolumn{1}{c|}{11.1G} & \multicolumn{1}{c|}{31.3G} & \multicolumn{1}{c|}{8.00G} & \CompCellColor{7.17G} \\ \cline{2-13} 
 & 7h & \ReduCellColor{69.2\%} & \ToolCellColor{3.60G} & 31.3G & 1.06G & 18.5G & 31.3G & \multicolumn{1}{c|}{31.3G} & \multicolumn{1}{c|}{17.5G} & \multicolumn{1}{c|}{31.3G} & \multicolumn{1}{c|}{13.6G} & \CompCellColor{11.7G} \\ \cline{2-13} 
 & 10h & \ReduCellColor{69.3\%} & \ToolCellColor{4.23G} & 31.3G & 1.41G & 22.2G & 31.3G & \multicolumn{1}{c|}{31.3G} & \multicolumn{1}{c|}{20.8G} & \multicolumn{1}{c|}{31.3G} & \multicolumn{1}{c|}{17.1G} & \CompCellColor{13.8G} \\ \hline
\multirow{3}{*}{tiffinfo} & 3h & \ReduCellColor{38.8\%} & \ToolCellColor{1.09G} & 2.22G & {934M} & 31.3G & \CompCellColor{1.78G} & \multicolumn{1}{c|}{2.75G} & \multicolumn{1}{c|}{31.3G} & \multicolumn{1}{c|}{31.3G} & \multicolumn{1}{c|}{31.3G} & 31.3G \\ \cline{2-13} 
 & 7h & \ReduCellColor{32.2\%} & \ToolCellColor{1.94G} & 4.08G & {1.60G} & 31.3G & \CompCellColor{2.86G} & \multicolumn{1}{c|}{4.11G} & \multicolumn{1}{c|}{31.3G} & \multicolumn{1}{c|}{31.3G} & \multicolumn{1}{c|}{31.3G} & 31.3G \\ \cline{2-13} 
 & 10h & \ReduCellColor{32.5\%} & \ToolCellColor{2.35G} & 4.60G & {2.06G} & 31.3G & \CompCellColor{3.48G} & \multicolumn{1}{c|}{5.27G} & \multicolumn{1}{c|}{31.3G} & \multicolumn{1}{c|}{31.3G} & \multicolumn{1}{c|}{31.3G} & 31.3G \\ \hline
\multirow{3}{*}{jasper} & 3h & \ReduCellColor{86.0\%} & \ToolCellColor{449M} & 12.8G & {587M} & 5.49G & 31.3G & \multicolumn{1}{c|}{27.1G} & \multicolumn{1}{c|}{4.97G} & \multicolumn{1}{c|}{\CompCellColor{3.20G}} & \multicolumn{1}{c|}{7.84G} & 4.48G \\ \cline{2-13} 
 & 7h & \ReduCellColor{80.3\%} & \ToolCellColor{1.01G} & 22.3G & {1.13G} & 8.77G & 31.3G & \multicolumn{1}{c|}{31.3G} & \multicolumn{1}{c|}{7.95G} & \multicolumn{1}{c|}{\CompCellColor{5.12G}} & \multicolumn{1}{c|}{12.5G} & 7.41G \\ \cline{2-13} 
 & 10h & \ReduCellColor{80.3\%} & \ToolCellColor{1.22G} & 24.6G & {1.51G} & 10.1G & 31.3G & \multicolumn{1}{c|}{31.3G} & \multicolumn{1}{c|}{9.54G} & \multicolumn{1}{c|}{\CompCellColor{6.15G}} & \multicolumn{1}{c|}{15.0G} & 8.56G \\ \hline
\multirow{3}{*}{transicc} & 3h & \ReduCellColor{64.9\%} & \ToolCellColor{6.10G} & 21.9G & {831M} & 31.3G & 31.3G & \multicolumn{1}{c|}{31.3G} & \multicolumn{1}{c|}{31.3G} & \multicolumn{1}{c|}{\CompCellColor{17.4G}} & \multicolumn{1}{c|}{31.3G} & 31.3G \\ \cline{2-13} 
 & 7h & \ReduCellColor{56.2\%} & \ToolCellColor{13.7G} & \CompCellColor{31.3G} & {1.00G} & 31.3G & 31.3G & \multicolumn{1}{c|}{31.3G} & \multicolumn{1}{c|}{31.3G} & \multicolumn{1}{c|}{31.3G} & \multicolumn{1}{c|}{31.3G} & 31.3G \\ \cline{2-13} 
 & 10h & \ReduCellColor{45.0\%} & \ToolCellColor{17.2G} & \CompCellColor{31.3G} & {1.14G} & 31.3G & 31.3G & \multicolumn{1}{c|}{31.3G} & \multicolumn{1}{c|}{31.3G} & \multicolumn{1}{c|}{31.3G} & \multicolumn{1}{c|}{31.3G} & 31.3G \\ \hline
\multirow{3}{*}{flvmeta} & 3h & \ReduCellColor{88.5\%} & \ToolCellColor{377M} & 3.42G & {560M} & 31.3G & \CompCellColor{3.27G} & \multicolumn{1}{c|}{4.34G} & \multicolumn{1}{c|}{12.6G} & \multicolumn{1}{c|}{31.3G} & \multicolumn{1}{c|}{15.4G} & 31.3G \\ \cline{2-13} 
 & 7h & \ReduCellColor{90.5\%} & \ToolCellColor{489M} & 10.8G & {911M} & 31.3G & \CompCellColor{5.15G} & \multicolumn{1}{c|}{7.01G} & \multicolumn{1}{c|}{17.2G} & \multicolumn{1}{c|}{31.3G} & \multicolumn{1}{c|}{25.7G} & 31.3G \\ \cline{2-13} 
 & 10h & \ReduCellColor{93.5\%} & \ToolCellColor{492M} & 11.7G & {1.16G} & 31.3G & \CompCellColor{7.57G} & \multicolumn{1}{c|}{7.97G} & \multicolumn{1}{c|}{19.8G} & \multicolumn{1}{c|}{31.3G} & \multicolumn{1}{c|}{31.3G} & 31.3G \\ \hline
\multirow{3}{*}{curl} & 3h & -85.2\% & 678M & 496M & 301M & 1.51G & 368M & \multicolumn{1}{c|}{366M} & \multicolumn{1}{c|}{31.3G} & \multicolumn{1}{c|}{26.1G} & \multicolumn{1}{c|}{5.22G} & 31.3G \\ \cline{2-13} 
 & 7h & -38.8\% & 680M & 578M & 339M & 3.24G & 490M & \multicolumn{1}{c|}{508M} & \multicolumn{1}{c|}{31.3G} & \multicolumn{1}{c|}{29.7G} & \multicolumn{1}{c|}{31.3G} & 31.3G \\ \cline{2-13} 
 & 10h & -15.0\% & 681M & 647M & 365M & 5.97G & 592M & \multicolumn{1}{c|}{641M} & \multicolumn{1}{c|}{31.3G} & \multicolumn{1}{c|}{28.8G} & \multicolumn{1}{c|}{31.3G} & 31.3G \\ \hline
\end{tabular}
    \label{tab:mem_comp}
\end{table*}

\subsubsection{Code Coverage}

KLEE merges the program bitcode with the bitcode of external libraries to produce a final complete bitcode before symbolic execution. In our evaluations, we collect coverage only for the source code in the program binary, excluding external libraries (e.g. C standard library). This is because external libraries are shared across programs, and their coverage statistics are not meaningful in this context. There are two options~\cite{coverage0000klee} for measuring code coverage for programs symbolically executed on KLEE. The first is \textbf{internal coverage}, which is reported directly by KLEE. A statement is considered covered if it has been symbolically executed. The second is \textbf{external coverage}, which is calculated by replaying the test inputs generated by KLEE on a native version of the program instrumented with \ProgramStyle{gcov}~\cite{gcov0000gcov} or \ProgramStyle{llvm-cov}~\cite{llvmcov0000cov}. We use both metrics to evaluate the code coverage of \ToolName{} and the baselines. We report internal coverage for basic blocks and external coverage for lines of code. This distinction aligns with \ToolName{}'s focus on basic blocks in CFGs, while replaying via \ProgramStyle{gcov} typically reports line coverage. However, due to the fact that \textsc{Learch} and \SearchToolStyle{cgs} are based on different versions of LLVM and KLEE, it is not possible to completely eliminate the inherent impact of version differences when reporting internal coverage. As a result, we report internal coverage of basic blocks for Group A experiments and external coverage of code lines for Group B experiments.

Figure~\ref{fig:block_cov} presents the experimental results for experiment Group A. Among the 12 benchmark programs, it is clear that \ToolName{} emerges as the best-performing search strategy in 10 of them. It also shows significant improvements in internal coverage for several programs, such as \ProgramStyle{bc}, \ProgramStyle{readelf}, and \ProgramStyle{tiffinfo}. For \ProgramStyle{bison}, \ToolName{} ranks second, trailing \SearchToolStyle{bfs} by 200 basic blocks. In the case of \ProgramStyle{make}, \ToolName{} ranks third, closely matching the coverage achieved by \SearchToolStyle{nurs:rp} and \SearchToolStyle{dfs}. Overall, \ToolName{} demonstrates consistently strong performance across the 12 benchmark programs, in contrast to the highly variable results exhibited by KLEE's baseline search strategies. The standard deviations in results for some programs are relatively high. This is because \ToolName{} explores fewer paths, which can lead to encountering complex path constraints more frequently. Some of these constraints are difficult for the SMT solver to resolve, occasionally resulting in a negative impact on \ToolName{}'s performance. Besides, our infeasible path handling includes some random choices for dependent states. Despite these challenges, \ToolName{} generally outperforms KLEE's baselines in covering basic blocks. In total, \ToolName{} covers 19853 basic blocks, whereas the best baseline search strategy, \SearchToolStyle{bfs}, covers 16602 basic blocks. This represents an overall improvement of 19.6\% compared to KLEE's baselines.

The experimental results for experiment Group B are largely consistent with those of Group A, as shown in Figure~\ref{fig:line_cov}. \ToolName{} achieves the best performance in 5 out of the 8 benchmark programs across all baseline search strategies, including prior works. In the case of \ProgramStyle{flvmeta}, \ToolName{} does not achieve the highest line coverage compared to its internal coverage. This is because certain paths that cover new lines are no longer executed when they do not align with the paths in the MPCs. Overall, \ToolName{} covers 17634 lines, while the best baseline search strategy, \SearchToolStyle{cgs}, covers 14173 lines across the 8 benchmark programs. This represents an overall improvement of 24.4\% compared to all baseline search strategies.

\subsubsection{Resource Usage}

In addition to measuring code coverage, we evaluate resource usage, including execution states and memory consumption, to demonstrate that our high-level approach is effective. We conduct experiments on both Group A and Group B. For Group A, memory usage can be easily obtained during execution using KLEE's memory manager, which allows us to track heap memory usage at any point in time. However, for Group B, comparing memory usage is not feasible because \textsc{Learch} and \SearchToolStyle{cgs} are based on older versions of KLEE with different internal implementation details. Under these circumstances, a comparison of memory usage would not be reliable. To address this limitation, we instead compare the number of execution states for Group B, as the definition of states remains consistent across different versions of KLEE.

Table~\ref{tab:mem_comp} presents the memory usage of \ToolName{} and KLEE baselines across 12 benchmark programs at the 3rd, 7th, and 10th hours of execution. We do not explicitly compare \ToolName{} with \SearchToolStyle{dfs}, as it is inherently optimized to minimize the number of states at the graph level. Due to its depth-first nature, \SearchToolStyle{dfs} maintains a small state pool, resulting in minimal memory consumption. However, this does not lead to higher code coverage for the majority of programs. Excluding \SearchToolStyle{dfs}, \ToolName{} achieves the lowest memory usage in 7 out of the 12 benchmark programs, reducing memory consumption by up to 93.5\% on \ProgramStyle{flvmeta}. Compared to other search strategies, \ToolName{} effectively reduces memory usage. Furthermore, we report the number of execution states in Table~\ref{tab:states_comp}. The results are consistent: \ToolName{} maintains a small number of states in 5 out of the 8 benchmark programs and reduces the number of execution states by up to 88.6\% on \ProgramStyle{jasper}.

\vspace{0.3cm}
\begin{longfbox}
\textbf{Result 1:} 
\ToolName{} increases basic block coverage by 19.6\% compared to KLEE search strategies and line coverage by 24.4\% compared to KLEE search strategies and 3 prior works. Moreover, \ToolName{} reduces memory usage by up to 93.5\% compared to KLEE search strategies and the number of execution states by up to 88.6\% compared to KLEE search strategies and 3 prior works.
\end{longfbox}

\subsection{RQ2: Finding Security Violations}

We evaluate the ability of all search strategies to discover security violations using two metrics. The first metric leverages KLEE's runtime Undefined Behavior Sanitizer (UBSan) support, and the second metric involves replaying test inputs on programs instrumented with sanitizers. \ProgramStyle{nasm} cannot be successfully compiled with UBSan, so it is excluded from the experiments in this section, leaving 11 benchmark programs for evaluation.

\begin{figure}[!]
\centering

\subfloat[Unsigned integer overflow at src/hash.c:397 in \ProgramStyle{make}]{
\includegraphics[width=0.22\textwidth]{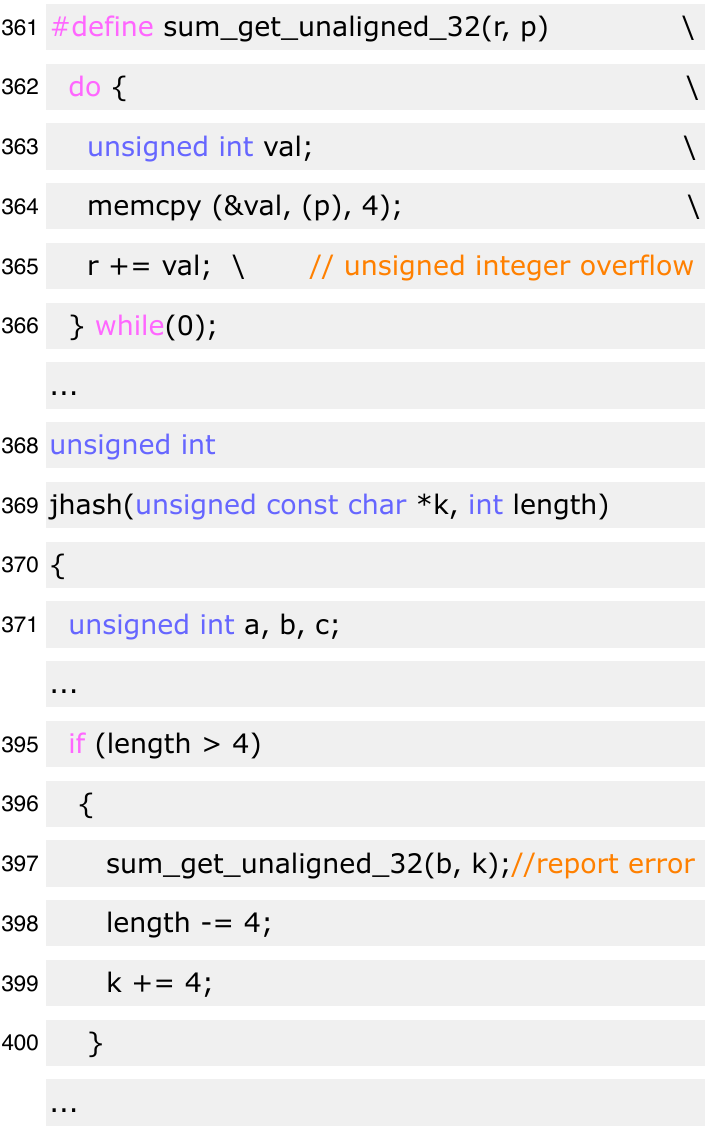}
\label{subfig:make_bug}
}
\hspace{0.01cm}
\subfloat[Implicit unsigned integer truncation at src/lcms2\_internal.h:184 in \ProgramStyle{transicc}]{
\includegraphics[width=0.22\textwidth]{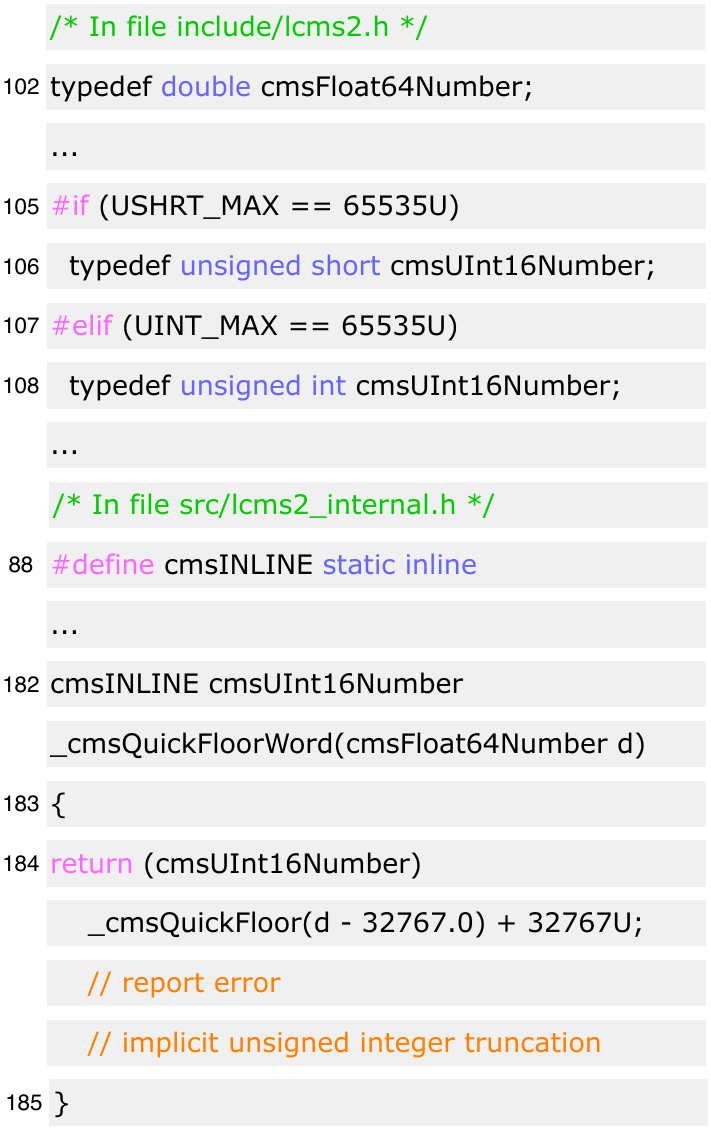}
\label{subfig:transicc_bug}}

\caption{\textbf{\small The two UBSan violations discovered by \ToolName{} based on KLEE UBSan support.}}
\label{fig:run_bugs}
\end{figure}

For the first metric, KLEE provides runtime support to detect certain UBSan bugs through its built-in functionality. However, since this support is limited, \ToolName{} identifies only two UBSan bugs, as shown in Figure~\ref{fig:run_bugs}, which is consistent with the results of KLEE's baseline search strategies, as they also detect the same two UBSan bugs.

\begin{figure}[!]
\centering

\subfloat[Memory leak in \ProgramStyle{bc}]{
\includegraphics[width=0.22\textwidth]{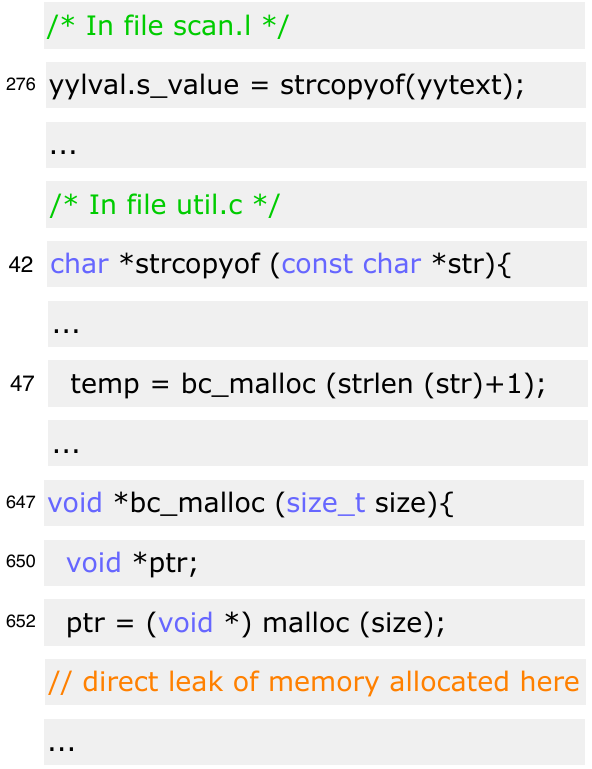}
\label{subfig:bc_bug}
}
\hspace{0.01cm}
\subfloat[Multiple undefined behaviors in \ProgramStyle{jasper}]{
\includegraphics[width=0.22\textwidth]{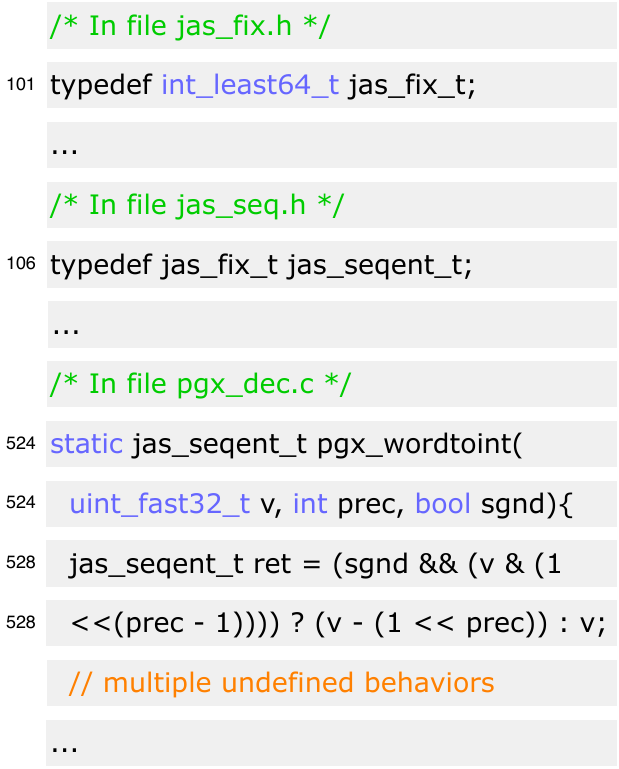}
\label{subfig:jasper_bug}}

\caption{\textbf{\small One ASan violation and one UBSan violation discovered by \ToolName{} by replaying test inputs.}}
\label{fig:replay_bugs}
\end{figure}

\begin{figure}[!t]
\centering
\includegraphics[scale=0.52]{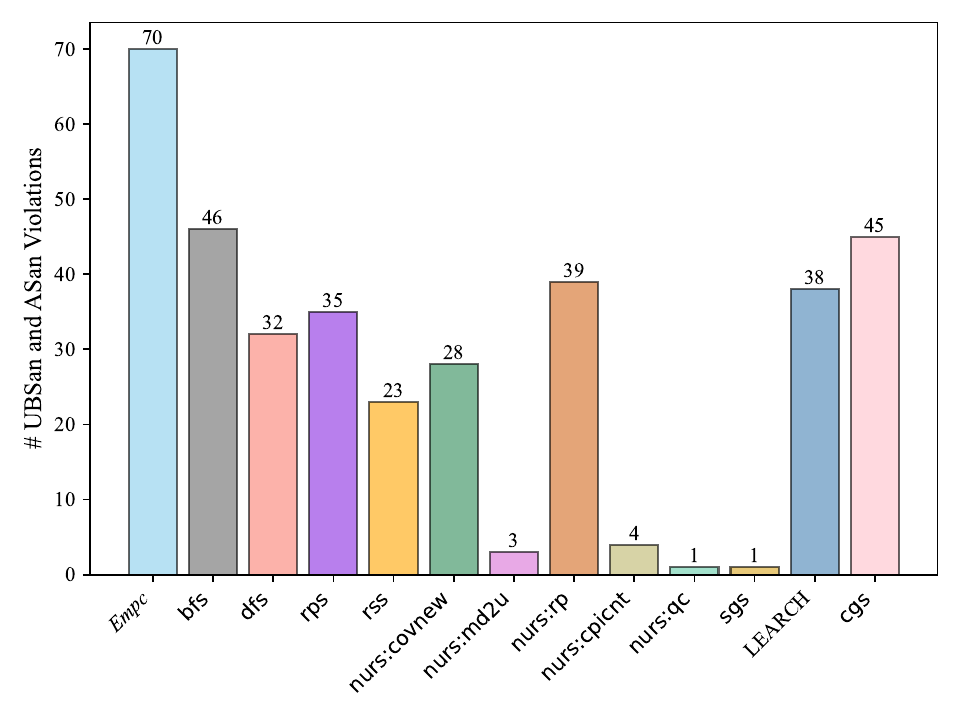}
\caption{\textbf{\small The total number of UBSan and ASan violations discovered by \ToolName{} and all baseline strategies via replay on 8 benchmark programs. This experiment is repeated 5 times and all security violations are collected.}}
\label{fig:bugs}
\end{figure}

Based on the second metric, we replay the test cases generated by all search strategies on programs instrumented with UBSan and Address Sanitizer (ASan) to identify additional security violations. This experiment is conducted on 8 benchmark programs (i.e., the 8 programs compatible with \textsc{Learch} and \SearchToolStyle{cgs}, as introduced in Section~\ref{sec:code_cov}). The total number of security violations discovered by all search strategies on these 8 programs is reported in Figure~\ref{fig:bugs}. \ToolName{} detects a total of 70 security violations, outperforming the second-best baseline, \SearchToolStyle{bfs}, by 24 violations. Furthermore, \ToolName{} uniquely identifies 15 new security violations that are not detected by any of the baseline search strategies. \revise{These results demonstrate that \ToolName{} is more effective at detecting security violations compared to baseline search strategies. This is because \ToolName{} prioritizes certain paths and maximizes code coverage within a limited time frame.}

We illustrate four examples of security violations found by \ToolName{} in Figure~\ref{fig:run_bugs} and Figure~\ref{fig:replay_bugs}. In Figure~\ref{subfig:make_bug}, the variable \CodeStyle{r} adds \CodeStyle{val} in a loop, resulting in an unsigned integer overflow. In Figure~\ref{subfig:transicc_bug}, the return value is truncated from 32 bits to 16 bits. These two UBSan violations are detected by \ToolName{} using KLEE's runtime UBSan support. Figure~\ref{subfig:bc_bug} demonstrates a memory leak: \CodeStyle{yylval} copies text using \CodeStyle{malloc} but does not release the allocated memory from the heap. Lastly, Figure~\ref{subfig:jasper_bug} highlights a common violation that involves integer conversion. The variable is implicitly converted from type \CodeStyle{int} to type \CodeStyle{unsigned long}, altering its value. Additionally, the variable is left-shifted by 31 places, causing an overflow as the result cannot be represented within the type \CodeStyle{int}.

Although UBSan and ASan violations, such as memory leaks, may not cause a program to crash, they can indicate potential issues or unexpected behavior in deeper parts of the programs. Since our experiments are conducted on the latest versions of real-world programs, we have reported the 70 discovered security violations to the respective developers and received confirmation of the existence of 16 ASan and UBSan violations.

\vspace{0.3cm}
\begin{longfbox}
\textbf{Result 2:} 
\ToolName{} is able to find more security violations than other search strategies on the 8 benchmark programs. It outperforms the second-best baseline search strategy by 24 violations.
\end{longfbox}

\subsection{RQ3: Runtime Overhead} \label{sec:overhead}

In this experiment, we evaluate the overhead introduced by \ToolName{} to the symbolic execution engine KLEE. \ToolName{} comprises two analysis components performed prior to symbolic execution, and we report the time taken by each component as well as the overall analysis time in Table~\ref{tab:analy_overhead}. The experiment is repeated 10 times, and the arithmetic mean is calculated. The results show that the analysis time for 10 out of 12 programs is limited to just a few dozen seconds. Even for more complex programs, such as \ProgramStyle{strip-new}, the analysis duration does not exceed 400 seconds. Consequently, the overall overhead introduced by \ToolName{} accounts for at most 1\% of the total execution time. This overhead is negligible compared to the total run time of the programs, indicating that the analysis performed by \ToolName{} has minimal impact on performance.

\begin{table}[!]
    \centering
    \caption{\textbf{\small The analysis performance overhead of \ToolName{} on all benchmark programs. It contains the time of graph analysis, dependence analysis and the overall analysis. The experiment is repeated 5 times and the mean is reported.}}
    \begin{tabular}[t]{lllr}
        \toprule
        \textbf{Program} & \textbf{Graph Analy.} & \textbf{Dep. Analy.} & \textbf{Overall} \\
        \midrule
         \ProgramStyle{bc} & 561ms & 262ms & 823ms \\
         \ProgramStyle{tic} & 11.9s & 1.5s & 13.4s \\
         \ProgramStyle{make} & 11.2s & 1.5s & 12.7s \\
         \ProgramStyle{bison} & 18.2s & 15.8s & 34.0s \\
         \ProgramStyle{readelf} & 46.9s & 28.9s & 75.8s \\
         \ProgramStyle{strip-new} & 178s & 200s & 378s \\
         \ProgramStyle{nasm} & 15.0s & 14.1s & 29.1s \\
         \ProgramStyle{tiffinfo} & 39.3s & 4.6s & 43.9s \\
         \ProgramStyle{jasper} & 25.3s & 2.3s & 27.6s \\
         \ProgramStyle{transicc} & 12.3s & 3.0s & 15.3s \\
         \ProgramStyle{flvmeta} & 12.6s & 1.1s & 13.7s \\
         \ProgramStyle{curl} & 57.6s & 81.8s & 139.4s \\
        \bottomrule
    \end{tabular}
    \label{tab:analy_overhead}
\end{table}

Since runtime path prioritization is integrated into the state selection and update methods in KLEE, we measure the runtime overhead associated with these operations. We run \ToolName{} and \SearchToolStyle{nurs:rp} on 12 benchmark programs for 10 hours, recording the total time spent on state selection and update during execution. This experiment is repeated 10 times, and the arithmetic mean is calculated. The comparison results are presented in Table~\ref{tab:runtime_overhead}. For 8 out of 12 programs, the total time spent on state selection remains less than 400 seconds, which is negligible compared to the overall run time of these programs. On average, \ToolName{} introduces a 12\% overhead for engine maintenance across all programs. However, for more complex programs, such as \ProgramStyle{transicc} and \ProgramStyle{nasm}, the time required for this procedure is longer. This increase in time consumption is attributed to \ToolName{}'s need to record, update, and manage all state operations during the selection and update process. Additionally, \ToolName{} performs more graph-level operations, further extending this phase. Despite the increased time required for these operations, it is important to emphasize that this does not adversely affect code coverage or memory usage.

\begin{table}[!]
    \centering
    \caption{\textbf{\small The runtime performance overhead of \ToolName{} and \SearchToolStyle{nurs:rp} on benchmark programs, including the time of state selection and state update handled by the searchers. The experiment is repeated 5 times and the mean is reported.}}
    \begin{tabular}[t]{lllr}
        \toprule
        \textbf{Program} & \textbf{\ToolName{}} & \textbf{\SearchToolStyle{nurs:rp}}  \\
        \midrule
         \ProgramStyle{bc} & 35ms & 5.8ms \\
         \ProgramStyle{tic} & 190min & 42.8s \\
         \ProgramStyle{make} & 122min & 83.5s \\
         \ProgramStyle{bison} & 1.9s & 36ms \\
         \ProgramStyle{readelf} & 270s & 2.7s \\
         \ProgramStyle{strip-new} & 267s & 721ms \\
         \ProgramStyle{nasm} & 283min & 61.5s \\
         \ProgramStyle{tiffinfo} & 279s & 5.7s \\
         \ProgramStyle{jasper} & 312s & 34s \\
         \ProgramStyle{transicc} & 321min & 6.5s \\
         \ProgramStyle{flvmeta} & 24.8s & 9.3s \\
         \ProgramStyle{curl} & 421s & 611ms \\
        \bottomrule
    \end{tabular}
    \label{tab:runtime_overhead}
\end{table}

\vspace{0.3cm}
\begin{longfbox}
\textbf{Result 3:} 
\ToolName{} adds at most 1\% overhead to the overall analysis. \ToolName{} adds at most 1\% overhead for symbolic execution engine maintenance on 8 out of 12 programs and adds 12\% overhead in average for engine maintenance on all programs.
\end{longfbox}

\subsection{RQ4: Design Choice} \label{sec:rq4}

\revise{
Firstly, we provide an experimental analysis of extraordinary loops in the 12 benchmark programs. Extraordinary loops, as described in Section \ref{sec:mpc_construct}, are loops with multiple loop headers. Among the 12 benchmark programs, we identify 68 extraordinary loops out of a total of 21568 loops, accounting for only 0.3\% of all loops. This result indicates that extraordinary loops are rare in real-world programs.
}

Secondly, we investigate the contributions of different components of \ToolName{}. \ToolName{} comprises two main types of analysis: graph analysis, which is used for the computation of MPCs, and dependence analysis, which handles infeasible paths. Since graph analysis is the fundamental high-level concept of \ToolName{}, we begin by evaluating the performance impact of dependence analysis. To do this, we remove dependence analysis from the implementation and replace the method for handling infeasible paths with a \SearchToolStyle{dfs} selection strategy. We then conduct experiments on 5 benchmark programs, repeating each experiment 5 times. Table \ref{tab:ablation} presents the comparison results between the original \ToolName{} and its modified version.


\revise{
In our ablation experiments, integrating dependence analysis into \ToolName{} generally results in an increase in code coverage for 11 out of 12 benchmark programs, with an average increase of 16\%. This demonstrates that program dependence analysis plays an important role in \ToolName{}, contributing to the observed code coverage improvements. However, the contribution of our MPC method is even more significant, as the increase introduced by program dependence analysis is only 16\%. Overall, these results suggest that the primary driver of the code coverage improvement is the MPC method, which provides the most substantial contribution. In contrast, dependence analysis plays a secondary, yet supportive, role in enhancing \ToolName{}'s performance.
}

\begin{table}[!]
    \centering
    \caption{\textbf{\small The final internal coverage of basic blocks of \ToolName{} and the modified version without dependence analysis. The experiment is repeated 5 times and the mean is reported.}}
    \begin{tabular}[t]{llr}
        \toprule
        \textbf{Program} & \textbf{\ToolName{}} & \textbf{Modified Version}\\
        \midrule
         \ProgramStyle{bc} & 395 & 236 \\
         \ProgramStyle{tic} & 915 & 860 \\
         \ProgramStyle{make} & 1863 & 1820 \\
         \ProgramStyle{bison} & 858 & 1100 \\
         \ProgramStyle{readelf} & 1683 & 1570 \\
         \ProgramStyle{strip-new} & 1930 & 1359 \\
         \ProgramStyle{nasm} & 1830 & 1815 \\
         \ProgramStyle{tiffinfo} & 1382 & 989 \\
         \ProgramStyle{jasper} & 2553 & 2518 \\
         \ProgramStyle{transicc} & 1828 & 1765 \\
         \ProgramStyle{flvmeta} & 1378 & 928 \\
         \ProgramStyle{curl} & 3233 & 3200 \\
        \bottomrule
    \end{tabular}
    \label{tab:ablation}
\end{table}


\vspace{0.3cm}
\begin{longfbox}
\textbf{Result 4:} 
Our results empirically support \ToolName{}'s design choices, demonstrating that the MPC method is the primary contributor, while program dependence plays a supportive role.
\end{longfbox}

\section{Limitations and Future Work}

\revise{
In this paper, we propose a novel approach that leverages minimum path covers to prioritize paths for symbolic execution, addressing the path explosion problem. To address the challenge of infeasible paths at run-time, we incorporate program dependence to refine path prioritization. In our evaluation, we compare \ToolName{} only with KLEE-based search strategies because \ToolName{} is implemented on KLEE. In the future, we plan to extend \ToolName{} to other symbolic execution engines, such as angr \cite{wang2017angr}, to evaluate its performance more broadly. Additionally, we plan to explore the application of \ToolName{} in the field of fuzzing, including using test inputs generated by \ToolName{} as fuzzing seeds to uncover more security vulnerabilities. We also plan to compare \ToolName{}'s effectiveness with existing fuzzing tools to assess its potential in improving software testing and vulnerability detection.
}

\section{Related Work}


\vspace{0.1cm}
\noindent
\textbf{Minimum path cover.} Minimum path cover (MPC) is a classic problem in graph theory. Dilworth\cite{dilworth1987decomposition} and Fulkerson\cite{fulkerson1956note} proved it's an NP-hard problem for general graphs but can be done in polynomial time for directed acyclic graphs. Many algorithms were proposed to compute MPC based on either maximum matching\cite{fulkerson1956note, ntafos1979path} or maximum flow\cite{ntafos1979path}. MPC has been applied to various fields such as 
software testing and programming languages\cite{bertolino1994automatic, bertolino1996many, wang1989generalized, kowaluk2008path}, distributed computing\cite{ikiz2006efficient} and bioinformatics\cite{rizzi2014complexity}.


\vspace{0.1cm}
\noindent
\textbf{Symbolic execution.} Symbolic execution is a powerful technique in software analysis that systematically explores the execution paths of a program by treating inputs as symbolic variables instead of concrete values~\cite{king1976symbolic, cadar2013symbolic, baldoni2018survey, cadar2008klee, wang2017angr}. This technique has enabled an increasing number of applications~\cite{chau2017symcerts, hernandez2017firmusb, chau2019analyzing, susag2022symbolic, schwerhoff2016advancing, coppa2017rethinking, hu2020automated, kuts2021towards, bucur2014prototyping}. Research in this field has focused mainly on improving the scalability and efficiency of symbolic execution tools, particularly in addressing challenges such as path explosion, where the number of execution paths grows exponentially with the size of the program~\cite{cadar2013symbolic, baldoni2018survey}.

\vspace{0.1cm}
\noindent
\revise{\textbf{Concolic execution.} Concolic execution combines the concrete and symbolic execution of the code under test to overcome the limitations of purely symbolic approaches\cite{godefroid2005dart,sen2007concolic,sen2005cute}. By leveraging concrete inputs to guide exploration while using symbolic constraints to systematically cover program paths, many tools\cite{godefroid2005dart,sen2005cute,yun2018qsym,chen2022symsan,hu2024marco,cha2019concolic,cha2018automatically,liu2020legion} have effectively detected vulnerabilities and generated comprehensive test suites. Although powerful, concolic execution still faces challenges with path explosion and complex constraints when analyzing large software systems.}

\vspace{0.1cm}
\noindent
\textbf{Searching strategies tackling path explosion.} Researchers have proposed various strategies to address the path explosion problem. One common approach involves using heuristics to prioritize and selectively explore paths that are more likely to uncover errors, thereby reducing the resource burden~\cite{cadar2008klee, li2013steering, zhou2022ferry,bessler2021metrinome}. Another technique, known as state merging, combines similar execution paths to reduce the total number of paths~\cite{kuznetsov2012efficient, hussein2023structural}. Furthermore, integrating symbolic execution with other program analysis techniques and modern machine learning methods has proven effective in pruning infeasible paths early~\cite{bugrara2013redundant, trabish2018chopped, cha2020making}. Additionally, advanced constraint-solving techniques aim to solve multiple constraints simultaneously, thereby reducing the burden on the SMT solver~\cite{ganesh2007decision, de2008z3, kapus2020pending}. Collectively, these methods contribute to mitigating the path explosion problem, making symbolic execution a more practical and scalable tool for software verification and security testing.

\section{Conclusion}

In this paper, we introduce a new approach \ToolName{} to tackle the path explosion problem in the symbolic execution technique based on minimum path cover (MPC). MPC provides an option for runtime path selection in order to use the least number of paths to maximize code coverage. This method not only increases code coverage but also reduces resource usage like memory usage. Moreover, we add program dependence analysis to handle some infeasible paths at run-time. We implement \ToolName{} on KLEE and show its effectiveness in increasing code coverage and reducing memory usage on 12 benchmark programs.


\ifCLASSOPTIONcompsoc
  \section*{Acknowledgements}
\else
  \section*{Acknowledgement}
\fi

We sincerely appreciate the anonymous reviewers and our shepherd for their valuable feedback and guidance. We also thank Yuchong Xie and Qiao Zhang for their helpful comments and discussions that improved the paper.

\bibliographystyle{plain}
\bibliography{miscellaneous/references}





\appendices

\section{Proof of Theorem \ref{thm:graph_split}} \label{app:graph_split}

\begin{table*}[!ht]
    \centering
    \caption{\textbf{\small The symbolic arguments in KLEE format for benchmark programs. These symbolic arguments are configured based on prior works and program-specific usage information.}}
    \begin{tabular}{lp{15.5cm}}
        \toprule
        \textbf{Program} & \textbf{KLEE Symbolic Environment} \\
        \midrule
         \ProgramStyle{bc} & \CommandStyle{--sym-stdin 20}  \\
         \ProgramStyle{tic} & \CommandStyle{--sym-args 0 2 8 A --sym-files 1 100} \\
         \ProgramStyle{make} & \CommandStyle{-n -f A --sym-files 1 40}  \\
         \ProgramStyle{bison} & \CommandStyle{--sym-args 0 2 2 A --sym-files 1 100}  \\
         \ProgramStyle{readelf} & \CommandStyle{-a A --sym-files 1 100} \\
         \ProgramStyle{strip-new} & \CommandStyle{--sym-args 0 2 8 A --sym-files 1 100} \\
         \ProgramStyle{nasm} & \CommandStyle{--sym-args 0 2 2 A --sym-files 1 100} \\
         \ProgramStyle{tiffinfo} & \CommandStyle{--sym-args 0 3 8 A --sym-files 1 300} \\
         \ProgramStyle{jasper} & \CommandStyle{--input A --output B --input-format --sym-arg 3 --output-format --sym-arg 3 --sym-args 0 3 15 --sym-files 2 300 --save-all-writes} \\
         \ProgramStyle{transicc} & \CommandStyle{--sym-args 0 2 4 A B --sym-files 2 200 --save-all-writes} \\
         \ProgramStyle{flvmeta} & \CommandStyle{--sym-arg 2 --sym-args 0 2 6 A --sym-files 1 300 --save-all-writes} \\
         \ProgramStyle{curl} & \CommandStyle{--sym-args 0 4 6 --sym-args 0 1 20} \\
        \bottomrule
    \end{tabular}
    \label{tab:bench_klee_sym}
\end{table*}

\begin{table*}[!ht]
    \centering
    \caption{\textbf{\small The number of execution states held by a KLEE instance is recorded at the 3rd, 7th, and 10th hours of execution. This evaluation is conducted only on programs that are compatible with all baseline search strategies. Reduction*: We exclude comparisons with \SearchToolSmallStyle{dfs} due to its specificity. We calculate the proportion of states reduced by \ToolName{} relative to the strategy with the minimum number of states. Non-negative reductions are highlighted in green cells.}}
    \scalebox{0.94}{
\begin{tabular}{|c|l|c|c|c|c|c|c|ccccc|c|c|c|}
\hline
\multirow{2}{*}{Program} & \multirow{2}{*}{Time} & \multirow{2}{*}{Reduction*} & \multirow{2}{*}{\ToolName{}} & \multirow{2}{*}{\SearchToolSmallStyle{bfs}} & \multirow{2}{*}{\SearchToolSmallStyle{dfs}*} & \multirow{2}{*}{\SearchToolSmallStyle{rss}} & \multirow{2}{*}{\SearchToolSmallStyle{rps}} & \multicolumn{5}{c|}{\SearchToolSmallStyle{nurs}} & \multirow{2}{*}{\SearchToolSmallStyle{sgs}} & \multirow{2}{*}{\textsc{Learch}} & \multirow{2}{*}{\SearchToolSmallStyle{cgs}} \\ \cline{9-13}
 &  &  &  &  &  &  &  & \multicolumn{1}{c|}{\SearchToolSmallStyle{rp}} & \multicolumn{1}{c|}{\SearchToolSmallStyle{covnew}} & \multicolumn{1}{c|}{\SearchToolSmallStyle{md2u}} & \multicolumn{1}{c|}{\SearchToolSmallStyle{cpicnt}} & \SearchToolSmallStyle{qc} &  &  &  \\ \hline
\multirow{3}{*}{bc} & 3h & -15.9\% & 175 & 355 & 151 & 173 & 287 & \multicolumn{1}{c|}{151} & \multicolumn{1}{c|}{255} & \multicolumn{1}{c|}{180K} & \multicolumn{1}{c|}{2.6K} & 132K & 453 & 492 & 220 \\ \cline{2-16} 
 & 7h & \ReduCellColor{34.4\%} & 217 & 772 & 141 & 432 & 528 & \multicolumn{1}{c|}{331} & \multicolumn{1}{c|}{812} & \multicolumn{1}{c|}{184K} & \multicolumn{1}{c|}{9.6K} & 222K & 1.2K & 1.4K & 1.4K \\ \cline{2-16} 
 & 10h & \ReduCellColor{31.6\%} & 310 & 1.4K & 155 & 717 & 633 & \multicolumn{1}{c|}{453} & \multicolumn{1}{c|}{1.3K} & \multicolumn{1}{c|}{187K} & \multicolumn{1}{c|}{16K} & 221K & 1.5K & 2.0K & 2.0K \\ \hline
\multirow{3}{*}{tic} & 3h & -935\% & 238K & 595K & 499 & 1.3M & 478K & \multicolumn{1}{c|}{569K} & \multicolumn{1}{c|}{1.3M} & \multicolumn{1}{c|}{1.7M} & \multicolumn{1}{c|}{409K} & 1.3M & 23K & 82K & 378K \\ \cline{2-16} 
 & 7h & -1153\% & 426K & 552K & 499 & 1.2M & 418K & \multicolumn{1}{c|}{516K} & \multicolumn{1}{c|}{1.6M} & \multicolumn{1}{c|}{1.7M} & \multicolumn{1}{c|}{360K} & 1.2M & 34K & 86K & 504K \\ \cline{2-16} 
 & 10h & -1166\% & 519K & 541K & 495 & 1.2M & 399K & \multicolumn{1}{c|}{505K} & \multicolumn{1}{c|}{1.6M} & \multicolumn{1}{c|}{1.7M} & \multicolumn{1}{c|}{360K} & 1.2M & 41K & 85K & 760K \\ \hline
\multirow{3}{*}{make} & 3h & -32.4\% & 4.9K & 575K & 61 & 544K & 576K & \multicolumn{1}{c|}{554K} & \multicolumn{1}{c|}{564K} & \multicolumn{1}{c|}{532K} & \multicolumn{1}{c|}{39K} & 556K & 13K & 3.7K & 111K \\ \cline{2-16} 
 & 7h & -79.2\% & 8.6K & 565K & 61 & 486K & 565K & \multicolumn{1}{c|}{549K} & \multicolumn{1}{c|}{529K} & \multicolumn{1}{c|}{440K} & \multicolumn{1}{c|}{127K} & 517K & 21K & 4.8K & 148K \\ \cline{2-16} 
 & 10h & -87.8\% & 9.2K & 560K & 61 & 457K & 562K & \multicolumn{1}{c|}{547K} & \multicolumn{1}{c|}{530K} & \multicolumn{1}{c|}{475K} & \multicolumn{1}{c|}{167K} & 481K & 26K & 4.9K & 175K \\ \hline
\multirow{3}{*}{bison} & 3h & -2.1\% & 146 & 772 & 70 & 2.5K & 512 & \multicolumn{1}{c|}{475} & \multicolumn{1}{c|}{3.1K} & \multicolumn{1}{c|}{3.0K} & \multicolumn{1}{c|}{143} & 2.6K & 2.1K & 497 & 335 \\ \cline{2-16} 
 & 7h & \ReduCellColor{14.0\%} & 208 & 1.1K & 94 & 3.3K & 760 & \multicolumn{1}{c|}{751} & \multicolumn{1}{c|}{3.9K} & \multicolumn{1}{c|}{3.7K} & \multicolumn{1}{c|}{242} & 3.7K & 2.7K & 921 & 525 \\ \cline{2-16} 
 & 10h & \ReduCellColor{34.3\%} & 222 & 1.3K & 105 & 3.6K & 860 & \multicolumn{1}{c|}{882} & \multicolumn{1}{c|}{4.3K} & \multicolumn{1}{c|}{4.4K} & \multicolumn{1}{c|}{338} & 4.5K & 3.3K & 1.3K & 700 \\ \hline
\multirow{3}{*}{tiffinfo} & 3h & -44.7\% & 6.8K & 53K & 241 & 1.6M & 44K & \multicolumn{1}{c|}{83K} & \multicolumn{1}{c|}{2.1M} & \multicolumn{1}{c|}{2.0M} & \multicolumn{1}{c|}{1.3M} & 1.6M & 31K & 74K & 4.7K \\ \cline{2-16} 
 & 7h & -60\% & 16K & 96K & 240 & 1.6M & 70K & \multicolumn{1}{c|}{133K} & \multicolumn{1}{c|}{2.1M} & \multicolumn{1}{c|}{2.1M} & \multicolumn{1}{c|}{1.4M} & 1.7M & 48K & 88K & 10K \\ \cline{2-16} 
 & 10h & -53.8\% & 20K & 105K & 222 & 1.6M & 85K & \multicolumn{1}{c|}{160K} & \multicolumn{1}{c|}{2.1M} & \multicolumn{1}{c|}{2.1M} & \multicolumn{1}{c|}{1.4M} & 1.6M & 58K & 93K & 13K \\ \hline
\multirow{3}{*}{jasper} & 3h & \ReduCellColor{88.6\%} & 3.3K & 632K & 27 & 115K & 1.4M & \multicolumn{1}{c|}{1.2M} & \multicolumn{1}{c|}{101K} & \multicolumn{1}{c|}{160K} & \multicolumn{1}{c|}{156K} & 128K & 29K & 90K & 350K \\ \cline{2-16} 
 & 7h & \ReduCellColor{58.1\%} & 18K & 1.1M & 27 & 182K & 1.5M & \multicolumn{1}{c|}{1.5M} & \multicolumn{1}{c|}{158K} & \multicolumn{1}{c|}{259K} & \multicolumn{1}{c|}{246K} & 208K & 43K & 90K & 914K \\ \cline{2-16} 
 & 10h & \ReduCellColor{62\%} & 19K & 1.2M & 27 & 218K & 1.5M & \multicolumn{1}{c|}{1.5M} & \multicolumn{1}{c|}{191K} & \multicolumn{1}{c|}{313K} & \multicolumn{1}{c|}{296K} & 221K & 50K & 93K & 1.2M \\ \hline
\multirow{3}{*}{transicc} & 3h & \ReduCellColor{62.1\%} & 7.2K & 57K & 1.5K & 1.6M & 832K & \multicolumn{1}{c|}{328K} & \multicolumn{1}{c|}{1.3M} & \multicolumn{1}{c|}{852K} & \multicolumn{1}{c|}{1.6M} & 1.6M & 19K & 53K & 38K \\ \cline{2-16} 
 & 7h & \ReduCellColor{40\%} & 15K & 67K & 1.7K & 1.4M & 1.1M & \multicolumn{1}{c|}{326K} & \multicolumn{1}{c|}{1.3M} & \multicolumn{1}{c|}{1.4M} & \multicolumn{1}{c|}{1.5M} & 1.5M & 25K & 124K & 73K \\ \cline{2-16} 
 & 10h & \ReduCellColor{37.9\%} & 18K & 70K & 1.9K & 1.4M & 1.1M & \multicolumn{1}{c|}{325K} & \multicolumn{1}{c|}{1.3M} & \multicolumn{1}{c|}{1.4M} & \multicolumn{1}{c|}{1.5M} & 1.5M & 29K & 120K & 82K \\ \hline
\multirow{3}{*}{flvmeta} & 3h & \ReduCellColor{78.1\%} & 4.6K & 192K & 264 & 1.4M & 104K & \multicolumn{1}{c|}{161K} & \multicolumn{1}{c|}{677K} & \multicolumn{1}{c|}{1.9M} & \multicolumn{1}{c|}{441K} & 1.5M & 21K & 141K & 2.9M \\ \cline{2-16} 
 & 7h & \ReduCellColor{81.9\%} & 5.8K & 646K & 272 & 1.4M & 163K & \multicolumn{1}{c|}{264K} & \multicolumn{1}{c|}{873K} & \multicolumn{1}{c|}{1.9M} & \multicolumn{1}{c|}{735K} & 1.4M & 32K & 62K & 4.5M \\ \cline{2-16} 
 & 10h & \ReduCellColor{83.7\%} & 6.2K & 673K & 262 & 1.4M & 199K & \multicolumn{1}{c|}{298K} & \multicolumn{1}{c|}{975K} & \multicolumn{1}{c|}{1.9M} & \multicolumn{1}{c|}{897K} & 1.4M & 38K & 95K & 4.9M \\ \hline
\end{tabular}
}
    \label{tab:states_comp}
\end{table*}

We first combine the MPCs in $G'$ and $G_{sub}$ to construct a path cover in $G$. If $k\geq |P_m^{sub}|$, we can easily expand $|P_m^{sub}|$ paths at $v_{sst}$ in $P'_{m}$ using the $|P_m^{sub}|$ paths in $P_m^{sub}$, and then the combined path cover 
$P_{com}$ in $G$ with $|P_{com}|=|P'_m|$; if $k < |P_m^{sub}|$, we can first expand $k$ paths at $v_{sst}$ in $P'_{m}$ using the $k$ paths in $P_m^{sub}$, and then add $|P_m^{sub}|-k$ paths in $P_m^{sub}$ and expand them into complete paths in $G$, so we can get the combined path cover $P_{com}$ in $G$ with $|P_{com}|=|P'_m|+|P_m^{sub}|-k$. Then we just need to prove $|P_{com}|=|P_m|$.

Assume the opposite of what we want to prove, which is $|P_{com}|\neq|P_m|$. There must be $|P_{com}|>|P_m|$ since $P_m$ is an MPC in $G$. Suppose there are $j$ paths going through $v_{ss}$ and $v_{st}$ in $P_m$. There must be $j\geq|P_m^{sub}|$ since $P_m^{sub}$ is an MPC in $G_{sub}$, so we can get $|P_m| \leq |P_m|-|P_m^{sub}| + j$. We remove the subgraph $G_{sub}$ and merge it as a vertex $v'_{sst}$ in the transformed graph $G'$. Let the paths in $P_m$ going through $v_{ss}$ and $v_{st}$ go through the merged vertex $v'_{sst}$, then we can get a new path cover $P''$ in $G'$ with $|P''|=|P_m|$. It's obvious that there are some identical paths in $P''$ due to the merged vertex $v'_{sst}$, and we can remove at least $l$ paths to make no identical paths appear in $P''$ with $0\leq l < j$. If we remove these $l$ paths, we can get a new path cover $P'''$ in $G'$ with $|P'''|=|P''|-l=|P_m|-l$. If $k\geq |P_m^{sub}|$, we have $|P_{com}|=|P'_m|$, so we can get $|P'''|<|P'_m|-l<|P'_m|$, which contradicts the assertion that $P'_m$ is an MPC in $G'$. If $k < |P_m^{sub}|$, we have $|P_{com}|=|P'_m|+|P_m^{sub}|-k$ and then we can get $|P'''|<|P'_m|+|P_m^{sub}|-k-l$. It's obvious that $j-l\leq k$ since $k$ is the maximum, so we have $k+l\geq j \geq |P_m^{sub}|$. Finally, we get $|P'''|<|P'_m|+|P_m^{sub}|-k-l < |P'_m|$, which also contradicts the assertion of MPC. Thus, there must be $|P_{com}|=|P_m|$.

\section{Proof of Theorem \ref{thm:loop_graph_split}} \label{app:loop_graph_split}

Similar to the proof in Appendix \ref{app:graph_split}, we can easily expand $|P_m^{sub}|$ paths at $v_{sst}$ and link vertices in $V_{st}$ in $P'_{m}$ using the $|P_m^{sub}|$ paths in $P_m^{sub}$. We still assume the opposite of what we want to prove, which is $|P_{com}|\neq|P_m|$. Thus, we conclude $|P_{com}|=|P_m|$.

\section{Proof of Theorem \ref{thm:mpc_matching}} \label{app:mpc_matching}

Before proving Theorem \ref{thm:mpc_matching}, we propose a lemma: for each MPC $P_m$, there must be at least one matching $M$ in $G_b$ that can be converted to $P_m$. This is because we can remove the disjoint vertices of these paths in $P_m$, and then we get multiple simple paths or isolated vertices, which can be represented in a matching $M$. Then assume the opposite of what we want to prove, that is, there is no maximum matching that can be converted to $P_m$. According to the lemma, we have a matching $M$ that can be converted to $P_m$. Ntaofs's work\cite{ntafos1979path} gives a theorem that provides a largest incomparable vertex set $|I_m|=|V|-|M_m|$. Thus, we cannot get a largest incomparable vertex set $I_m$ via matchings, which contradicts the assertion of $|I_m|=|P_m|$. Therefore, there must be a maximum matching $M_m$ corresponding to $P_m$.

\section{Symbolic Environment for Benchmarks} \label{app:bench_args}

The KLEE symbolic environment settings in our evaluations are listed in Table \ref{tab:bench_klee_sym}.

\section{Evaluation of Execution States} \label{app:states_num}

The evaluation results of the number of execution states are listed in Table \ref{tab:states_comp}.

\newpage

\section{Meta-Review}

The following meta-review was prepared by the program committee for the 2025
IEEE Symposium on Security and Privacy (S\&P) as part of the review process as
detailed in the call for papers.

\subsection{Summary}
This paper introduces Empc, a symbolic execution approach to address path explosion by modeling path selection as a minimum path cover problem on the program’s inter-procedural control flow graph. When a path in the cover is deemed infeasible, Empc dynamically adjusts and selects alternate paths. Implemented using KLEE, Empc boosts coverage by about 20\% (basic blocks) and 24.4\% (source lines), while reducing memory and symbolic state usage by up to 93.5\% and 88.6\%, respectively.

\subsection{Scientific Contributions}
\begin{itemize}
    \item Creates a New Tool to Enable Future Science.
    \item Provides a Valuable Step Forward in an Established Field.
\end{itemize}

\subsection{Reasons for Acceptance}
\begin{enumerate}
    \item The Program Committee appreciated the use of Path Cover to push forward the state of path prioritization, as this provides a good theoretical grounding compared to current techniques, which tend to rely on heuristics.
    \item The paper's commitment to open science and the release of a prototype was viewed as a significant contribution.
\end{enumerate}




\end{document}